\documentclass[prd,twocolumn,showpacs,preprintnumbers,floats,floatfix,
nofootinbib]{revtex4} 
\usepackage{graphicx} 
\usepackage{dcolumn} 
\usepackage{bm} 
\usepackage{amssymb} 
\def\spose#1{\hbox to 0pt{#1\hss}} 
\def\lta{\mathrel{\spose{\lower 3pt\hbox{$\mathchar"218$}} 
     \raise 2.0pt\hbox{$\mathchar"13C$}}} 
\def\gta{\mathrel{\spose{\lower 3pt\hbox{$\mathchar"218$}} 
     \raise 2.0pt\hbox{$\mathchar"13E$}}} 
\newcommand{\be}{\begin{equation}} 
\newcommand{\en}{\end{equation}} 
\newcommand{\bea}{\begin{eqnarray}} 
\newcommand{\ena}{\end{eqnarray}}

\def\setR{\mathbb{R}} 
 
\begin{document} 
 
\title{On the Dependence of the Spectra of Fluctuations in 
Inflationary Cosmology on Trans-Planckian Physics} 
 
\author{J\'er\^ome Martin}  
\email{jmartin@iap.fr} 
\affiliation{Institut d'Astrophysique de Paris-GReC0, 
98bis boulevard Arago, 
75014 Paris, France}  
 
\author{Robert Brandenberger}  
\email{rhb@het.brown.edu} 
\affiliation{Physics Department, Brown University, Providence, 
R.I. 02912 USA} 
 
\date{\today}  
 
\begin{abstract} 
We calculate the power spectrum of metric fluctuations in inflationary 
cosmology starting with initial conditions which are imposed mode by 
mode when the wavelength equals some critical length $\ell_{_{\rm C}}$ 
corresponding to a new energy scale $M_{_{\rm C}}$ at which 
trans-Planckian physics becomes important. In this case, the power 
spectrum can differ from what is calculated in the usual framework 
(which amounts to choosing the adiabatic vacuum state). The fractional 
difference in the results depends on the ratio $\sigma _0$ between the 
Hubble expansion rate $H_{\rm inf}$ during inflation and the new 
energy scale $M_{_{\rm C}}$. We show how and why different choices of 
the initial vacuum state (stemming from different assumptions about 
trans-Planckian physics) lead to fractional differences which depend 
on different powers of $\sigma _0$. As we emphasize, the power 
in general also depends on whether one is calculating the power 
spectrum of density fluctuations or of gravitational waves. 
\end{abstract} 
 
\maketitle 
 
\section{Introduction}

The exponential expansion of space in inflationary cosmology leads to 
the stretching of modes which were in the trans-Planckian regime at 
the beginning of inflation into the observable range. This leads to 
the possibility, first raised explicitly in \cite{RHBrev}, that 
trans-Planckian physics might be observable today in the cosmic 
microwave background. In earlier work \cite{MB1,BM2} we addressed this 
issue in a simple toy model obtained by replacing the linear 
dispersion relation of the cosmological fluctuations by new dispersion 
relations which differ from the linear one on length scales smaller 
than the Planck length (the same dispersion relations had been used 
earlier \cite{Unruh,CJ} in the context of an analysis of possible 
trans-Planckian effects on black hole radiation). We were able to 
construct dispersion relations which give rise to large (order one) 
corrections to the usual spectrum of fluctuations, but the price to pay 
is a fine-tuning of the parameters describing the model and/or a 
back-reaction problem. This question has been further analyzed in many 
papers (see for instance Refs.~\cite{Niemeyer,MB3,NP2,LLMU,BJM,S}). It 
was found that in order to obtain measurable differences in the 
predictions, non-adiabatic evolution of the state on trans-Planckian 
scales is required. 
 
\par 
 
In another line of approach to the {\it trans-Planckian challenge} to 
inflationary cosmology, the possibility of measurable effects of 
trans-Planckian physics on observables such as CMB anisotropies and 
power spectra of scalar and tensor metric fluctuations was studied 
\cite{kempf,Chu,Easther,kempfN,Hui,HoB,Hassan} in models where the 
trans-Planckian physics is based on stringy space-time uncertainty 
relations. In particular, the authors of \cite{Easther} found a 
spectrum with oscillations of amplitude $\sigma _0\equiv H_{\rm inf}/ 
M_{_{\rm C}}$, where $H_{_{\rm inf}}$ is the Hubble parameter during 
inflation and $M_{_{\rm C}}$ a characteristic scale at which the 
trans-Planckian physics shows up, superimposed on the usual 
scale-invariant spectrum, whereas the authors of \cite{kempfN} found 
only much smaller effects. 
 
\par 
 
The trans-Planckian problem was also tackled in the framework of 
non-commutative geometry in Ref.~\cite{LMMP}. It was found that the 
effect is of order $(H_{\rm inf}/ M_{_{\rm C}})^4$. It was also shown 
in this article that non-commutative geometry implies the presence of 
a preferred direction which would result in a correlation between 
different multipoles $C_{\ell }$ and $C_{\ell \pm2}$. 
 
\par 
 
In yet another approach to the trans-Planckian issue, Danielsson
\cite{Dan1} (see also Ref.~\cite{Lowe1}) suggested to replace the
unknown physics on trans-Planckian scales by assuming that the modes
representing cosmological fluctuations are generated mode by mode at
the time when the wavelength of the mode equals the Planck length, or
more generally when it equals the length $\ell_{_{\rm C}}$ associated
with the energy scale $M_{_{\rm C}}$ of the new physics which sets the
initial conditions.  There is a one-parameter family of vacuum states
($\alpha$ vacua) of a free quantum field in de Sitter space which can
be considered, and for nontrivial $\alpha$ vacua Danielsson found
effects of the choice of the initial state which are of linear order
in the ratio $\sigma _0$\footnote{Recently, Ref.~\cite{Alberghi} has
shown that effects of the order $\sigma _0$ also occur in models of
power-law inflation.}, and such effects could be seen in observations
\cite{Dan3}
\footnote{There has been a concern that nontrivial $\alpha$ vacua are 
problematic from the point of view of interacting quantum field theory 
\cite{Banks,Einhorn,Kaloper2}. However, very recently it has been 
shown \cite{Lowe2} how to define interacting quantum field theory 
about an $\alpha$ vacuum.}. Similar results were found by Easther et 
al. \cite{Easther3}, whereas Niemeyer et al. \cite{npc} have argued 
that if the modes are set off in the initial adiabatic vacuum state 
when their wavelength equals $\ell_{_{\rm C}}$, then the effects are 
of order $\sigma _0^3$ and hence (in usual models) completely 
negligible.   
 
Using an effective field theory method, Kaloper et 
al. \cite{Kaloper1} have argued that the effects of trans-Planckian 
physics on cosmological fluctuations should be at most of the order 
$\sigma _0^2$, assuming that the modes are in the adiabatic vacuum 
state when the wavelength is equal to the Hubble radius (see 
Ref.~\cite{BM4} for a criticism of imposing initial conditions at 
Hubble radius crossing, and see Ref.~\cite{Cline} for counterexamples 
to the claims of Ref.~\cite{Kaloper1}).   
 
In this paper, we re-consider 
the calculation of the spectrum of cosmological perturbation in the 
{\it minimal trans-Planckian} setting \cite{Dan1} when mode by mode 
the initial conditions for the mode are set when the wavelength equals 
the Planck length (or, more generally, the length scale of the new 
physics). We find that the overall amplitude of the correction terms (compared 
to the usual spectra) depends sensitively on the precise 
prescription of the initial state, it depends on whether 
one is studying power-law or slow-roll inflation, and it also depends 
on whether one is computing the spectrum of scalar metric fluctuations 
or of gravitational waves. Some of the ``discrepancies'' between 
the results of previous studies is due to the fact that 
different quantities were calculated in different models. 
We show that when the initial state is chosen to be the 
instantaneous Minkowski vacuum, then the deviations of the power 
spectrum from the usual result are of the order $\sigma _0^3$, in 
agreement with what was found in \cite{npc}. In an arbitrary $\alpha-$ 
vacuum, the choice of the value of $\alpha$ has an effect on the 
amplitude of the fluctuation spectrum which is not suppressed by any 
power of $\sigma _0$. However, if $\alpha$ is independent of $k$, the 
effect will not be distinguishable from a slight change in the 
underlying parameters of the background inflationary model. However, 
in general (and specifically in the choice of the vacuum made in 
\cite{Dan1}, the amplitude of the correction term in the power 
spectrum will have a k-dependent (and hence observable) piece which is 
first order in $\sigma _0$, at least in the case of the spectrum 
of gravitational waves. 

\par
 
While this paper was being finalized, three preprints appeared 
which investigate related aspects of the trans-Planckian problem. 
In Ref.~\cite{Venez}, the choice of various initial states was 
related to the minimization of different Hamiltonians. In Ref.~\cite{Armen}, 
the predictions of inflation for the power spectrum of fluctuations 
was studied for a two parameter class of initial states, and the 
amplitude of the corrections compared to the usual results was seen 
to depend sensitively on which state is chosen. In Ref.~\cite{Chung},
the fact that the definitions of the Bunch-Davies and of the adiabatic
vacua have some intrinsic ambiguities in a Universe with a de
Sitter phase of finite duration was analyzed.

\section{Cosmological perturbations of quantum-mechanical origin} 
 
\subsection{General considerations} 
 
The line element for the spatially flat 
Friedmann-Lemaitre-Robertson-Walker (FLRW) background plus small 
perturbations can be written as \cite{MFB}: 
\begin{eqnarray} 
\label{metricgi} 
{\rm d}s^2 &=&  
a^2(\eta )\{-(1-2\phi ){\rm d}\eta ^2+2({\rm \partial}_iB){\rm d}x^i 
{\rm d}\eta \nonumber \\
&+& [(1-2\psi )\delta _{ij}
+2{\rm \partial }_i{\rm \partial }_jE+h_{ij}]{\rm d}x^i{\rm d}x^j\}\, , 
\end{eqnarray} 
where the functions $\phi $, $B$, $\psi $ and $E$ represent the scalar 
sector whereas the traceless and transverse tensor $h_{ij}$ 
($h_i{}^i=h_{ij}{}^{,j}=0$), represents the tensor sector, i.e. the 
gravitational waves. The time $\eta $ is the conformal time and is 
related to the cosmic time $t$ by the relation ${\rm d}t=a(\eta ){\rm 
d}\eta $. It is convenient to introduce the background quantity 
$\gamma (\eta )$ defined by $\gamma \equiv -\dot{H}/H^2$, where a dot 
means differentiation with respect to cosmic time and $H$ is the 
Hubble rate, $H\equiv \dot{a}/a$. Using the conformal time we may 
rewrite $\gamma $ as $\gamma =1-{\cal H}'/{\cal H}^2$, where ${\cal 
H}\equiv a'/a$, and a prime denotes differentiation with respect to 
the conformal time. For power-law scale factors, $a(\eta )\propto 
(-\eta )^q$, the function $\gamma (\eta )$ turns out to be a constant, 
$\gamma =(q+1)/q$, which vanishes in the particular case of the de 
Sitter space-time characterized by $q=-1$. The perturbed Einstein 
equations provide us with the equations of motions for the 
cosmological perturbations. At the linear level, each type of 
perturbations decouple and we can treat them separately. 
 
\par 
 
In the tensor sector (which is automatically gauge-invariant) we 
define the quantity $\mu _{_{\rm T}}$ for each mode $k$ according to 
\begin{equation} \label{eq2} 
h_{ij}(\eta ,{\bf x})=\frac{1}{a}\frac{1}{(2\pi )^{3/2}} 
\sum _{s=1}^2\int {\rm d}{\bf k}p_{ij}^s({\bf k}) 
\mu _{_{\rm T}}(\eta ,{\bf k}){\rm e}^{i{\bf k}\cdot {\bf x}}\, , 
\end{equation} 
where $p_{ij}^s({\bf k})$ is the polarization tensor. The plane waves 
appear in the previous expression because the space-like sections are 
flat. At linear order, gravitational waves do not couple to matter 
and, as a consequence, the equation of motion is 
just given by the perturbed vacuum Einstein equation. Explicitly, it 
reads \cite{Grigw}: 
\begin{equation} 
\label{eomtensor} 
\mu _{_{\rm T}}''+\biggl(k^2-\frac{a''}{a}\biggr)\mu _{_{\rm T}}=0\ . 
\end{equation}  
This equation is similar to a time-independent Schr\"odinger equation
with an effective potential given by $U_{_{\rm T}}(\eta )=a''/a$. It
can also be viewed as the equation of a parametric oscillator whose
time-dependent frequency is given by $\omega _{_{\rm T}}=k^2-a''/a$,
see Ref.~\cite{MSwkb}. In the vacuum state, the two point correlation
function of gravitational waves reads
\begin{equation} 
\langle 0\vert h_{ij}(\eta ,{\bf x})h^{ij}(\eta ,{\bf x}+{\bf r}) 
\vert 0 \rangle =\int _0^{+\infty }\frac{{\rm d}k}{k}\frac{\sin kr}{kr} 
k^3P_h(k)\, . 
\end{equation} 
The power spectrum $k^3P_h(k)$ which appears in the expression for the  
two point correlation function is given by 
\begin{equation} 
k^3P_h(k)=\frac{2k^3}{\pi ^2}\biggl \vert \frac{\mu _{_{\rm T}}}{a} 
\biggr \vert ^2\, . 
\end{equation} 
This quantity is {\it a priori} time and wavenumber dependent but for 
super-horizon modes, it turns out to be time-independent because the  
growing mode is given by $\mu _{_{\rm T}}\propto a(\eta )$. 
 
\par 
 
Let us now turn to scalar metric (density) perturbations. The two most 
important differences with the gravitational waves are that the scalar 
sector is gauge-dependent and that the scalar perturbations of the 
metric are coupled to the perturbations of the stress-energy tensor 
describing the matter. Scalar perturbations of the geometry can be 
characterized by the two gauge-invariant Bardeen potentials $\Phi 
\equiv \phi +(1/a)[(B-E')a]'$ and $\Psi \equiv \psi -{\cal H}(B-E')$ 
\cite{Bardeen}. From now on, we restrict ourselves to the case where 
the matter is described by a scalar field: $\varphi=\varphi _0(\eta 
)+\varphi _1(\eta ,{\bf x})$. Fluctuations in the scalar field are 
characterized by the gauge-invariant quantity ${\rm \delta}\varphi 
\equiv \varphi _1+\varphi _0'(B-E')$. The full set of the scalar 
perturbed (gauge-invariant) Einstein equations is 
\begin{widetext} 
\begin{eqnarray} 
& & -3{\cal H}({\cal H}\Phi +\Psi ')+\partial _k\partial ^k\Psi 
=\frac{\kappa }{2} \biggl[-(\varphi _0')^2\Phi +\varphi _0'\delta 
\varphi ' +a^2\frac{{\rm d}V}{{\rm d}\varphi _0}\delta \varphi 
\biggr],  
\quad  
\partial _i({\cal H}\Phi +\Psi ' )=\frac{\kappa }{2}\varphi 
_0'\partial _i\delta \varphi \, ,  
\\  
& & \partial _i\partial ^j(\Phi -\Psi )=0\, , \quad (2{\cal 
H}'+{\cal H}^2)\Phi +{\cal H}\Phi '+\Psi ''+2{\cal H}\Psi ' 
-\frac{1}{3}\partial _k\partial ^k(\Phi -\Psi ) 
=\frac{\kappa }{2} \biggl[-(\varphi 
_0')^2\Phi +\varphi _0'\delta \varphi ' -a^2\frac{{\rm d}V}{{\rm 
d}\varphi _0}\delta \varphi \biggr]\, , 
\end{eqnarray} 
\end{widetext} 
where $\kappa =8\pi/m_{_{\rm Pl}}^2$, $m_{_{\rm Pl}}$ being the Planck 
mass. At this point, one has to be careful because the case $\varphi 
_0'=0$, which corresponds to the de Sitter space-time, is 
particular. In this situation, the solution of the above system of 
equations is simply $\Phi =0$, i.e. there are no density perturbations 
at all (it is also necessary to require that the Bardeen potential is 
finite at infinity). If $\varphi _0'\neq 0$, then everything can be 
reduced to the study of a single gauge-invariant variable (the 
so-called Mukhanov-Sasaki variable) defined by \cite{Mukh} 
\begin{equation} 
v\equiv a\biggl(\delta \varphi+ 
\frac{\varphi _0'}{{\cal H}}\Phi \biggr)\, . 
\end{equation} 
It turns out to be more convenient to work with the variable $\mu 
_{_{\rm S}}$ defined by $\mu _{_{\rm S}}\equiv -\sqrt{2\kappa }v$. Its 
equation of motion is very similar to that of the gravitational waves 
and reads \cite{MS}: 
\begin{equation} 
\label{eomscalar} 
\mu _{_{\rm S}}''+\biggl[k^2- 
\frac{(a\sqrt{\gamma})''}{(a\sqrt{\gamma })}\biggr]\mu _{_{\rm S}}=0\ . 
\end{equation} 
Some remarks are in order here. Firstly, let us stress again that the 
above equation can be used only if $\varphi _0'\neq 0$ or $\gamma \neq 
0$ because in course of its derivation, we have divided by $\varphi 
_0'$. Secondly, the effective potential for density perturbations, 
$U_{_{\rm S}}=(a\sqrt{\gamma })''/(a\sqrt{\gamma })$ is different from 
the potential for gravitational waves. Equivalently, its 
time-dependent frequency is also different and given by $\omega 
_{_{\rm S}}=k^2 -(a\sqrt{\gamma })''/(a\sqrt{\gamma })$. Thirdly, in 
the case of power-law inflation, the function $\gamma $ is a constant 
and we have $U_{_{\rm S}}=U_{_{\rm T}}$. Fourthly, density 
perturbations are often characterized by the so-called conserved 
quantity $\zeta $ defined by $\zeta \equiv (2/3)({\cal H}^{-1}\Phi 
'+\Phi )/(1+\omega )+\Phi $, where $\omega $ is the equation of state 
parameter, i.e. the ratio of pressure to energy density of the 
background scalar field. This quantity is directly linked to $\mu 
_{_{\rm S}}$ through the equation $\mu _{_{\rm S}}=-2a\sqrt{\gamma 
}\zeta $. The integration of Eq.~(\ref{eomscalar}) leads to the 
primordial spectrum of this quantity, namely 
\begin{equation} 
\label{specdp} 
k^3P_{\zeta }(k)=\frac{k^3}{8\pi ^2}\biggl\vert \frac{\mu 
_{_{\rm S}}}{a\sqrt{\gamma }}\biggr\vert ^2 \, , 
\end{equation} 
The fact that this quantity is meaningless for the de Sitter case, 
$\gamma =0$, is obvious. 
 
\par 
 
Let us now consider a different situation: there are no cosmological 
fluctuations anymore but just a test scalar field $\chi (\eta ,{\bf 
x})$ living in a FLRW universe. If we Fourier expand the scalar field 
according to 
\begin{equation} 
\chi (\eta ,{\bf x})=\frac{1}{a(\eta )}\frac{1}{(2\pi )^{3/2}} \int {\rm 
d}{\bf k}\mu (\eta, {\bf k}){\rm e}^{i{\bf k}\cdot {\bf x}}\, , 
\end{equation} 
then the Klein-Gordon equation takes the form 
\begin{equation} 
\label{motsf} 
\mu ''+\biggl(k^2-\frac{a''}{a}\biggr)\mu =0 \, . 
\end{equation} 
We recognize the equation of motion of gravitational waves. The two-point  
correlation function of the scalar field can be written as  
\begin{equation} 
\langle 0\vert \chi (\eta ,{\bf x})\chi (\eta ,{\bf x}+{\bf r}) 
\vert 0 \rangle =\int _0^{+\infty }\frac{{\rm d}k}{k}\frac{\sin kr}{kr} 
k^3P_{\chi }(k)\, . 
\end{equation} 
The quantity $k^3P_{\chi }(k)$ which appears in the expression for 
the two point correlation function is the power spectrum of the scalar 
field and is given by 
\begin{equation}
\label{specscaf} 
k^3P_{\chi }(k)=\frac{k^3}{2\pi ^2}\biggl \vert \frac{\mu }{a} 
\biggr \vert ^2\, . 
\end{equation} 
We also see that the equation of motion for density perturbations is
equivalent to the equation for a scalar field if the function $\gamma
$ is a non-vanishing constant. This is why, for this class of models,
the study of density perturbations is in fact equivalent to the study
of a scalar field. However, there is an important exception to this
prescription: the de Sitter space-time since, in this case, $\gamma
=0$ so that the equation of motion is not given by
Eq.~(\ref{eomscalar}) as already mentioned above. Therefore, it is
inconsistent to first assume that $\gamma $ is a constant, then to use
Eq.~(\ref{specdp}) in order to calculate the power spectrum and
finally to particularize the result to the de Sitter case. This
procedure can only lead to the determination of the spectrum of
gravitational waves and/or of a test scalar field but not of density
perturbations in the de Sitter space-time.
 
\subsection{Power-law inflation} 
 
The case of power-law inflation, for which the scale factor can be 
written as $a(\eta )=\ell _0(-\eta )^q$, is important because the 
equation of motion for the cosmological perturbations can be solved 
explicitly. The parameter $\ell _0$ is a length since we have chosen 
to work with a dimension-full scale factor. Since the function $\gamma 
(\eta )$ reduces to a constant, the effective potential for density 
perturbations simplifies to $a''/a$ and becomes identical to the 
gravitational waves effective potential. Its explicit form reads 
$U_{_{\rm S}}(\eta )=U_{_{\rm T}}(\eta )=q(q-1)/\eta ^2$. Then, the 
exact solution for the variables $\mu _{_{\rm S}}$, $\mu _{_{\rm T}}$ 
and $\mu $ reads 
\begin{equation} 
\mu _{_{\rm S,T}}=(k\eta )^{1/2} 
[A_1(k)J_{q-1/2}(k\eta )+A_2(k)J_{-q+1/2}(k\eta )]\, . 
\end{equation} 
In the above expression, $J_{\nu }$ is a Bessel function of order $\nu 
$. The two $k$-dependent constants $A_1(k)$ and $A_2(k)$ are fixed by the 
initial conditions. 

\par

We now consider the standard calculation of inflationary cosmology. In
this case, the initial conditions are fixed in the infinite past. When
$k\eta \rightarrow -\infty $, the mode function tends to a plane wave
with positive and negative frequency. The usual procedure requires
that
\begin{equation} 
\label{stic} 
\lim _{k/(aH)\rightarrow +\infty }\mu _{_{\rm S,T}}=\mp 
\frac{4\sqrt{\pi }}{m_{_{\rm Pl}}}
\frac{{\rm e}^{-ik\eta }}{\sqrt{2k}}\, . 
\end{equation} 
Let us notice that the origin of the minus sign for density 
perturbations is the minus sign in the relation $\mu _{_{\rm 
S}}=-\sqrt{2\kappa }v$, $v$ being the quantity which is canonically 
quantized, i.e. which behaves as ${\rm e}^{-ik\eta }/\sqrt{2k}$ in the 
ultraviolet limit. Then, the constants $A_1(k)$ and $A_2(k)$ are 
completely specified and are given by 
\begin{eqnarray} 
\frac{A_1(k)}{A_2(k) } &=& -{\rm e}^{i\pi (q-1/2)}\, , \\  
A_2(k) &=&  \mp \frac{2\pi }{m_{_{\rm Pl}}}\frac{{\rm e}^{-i\pi (q-1)/2}} 
{\cos (\pi q)}k^{-1/2}\, . 
\end{eqnarray} 
This implies that the mode functions $\mu _{_{\rm S,T}}$ can be 
expressed as 
\begin{equation} 
\mu _{_{\rm S,T}}^{\rm stand}(k,\eta )=\mp  
\frac{2i\pi }{m_{_{\rm Pl}}}(-\eta )^{1/2} 
{\rm e}^{-i\pi q/2}H^{(1)}_{1/2-q}(-k\eta )\, , 
\end{equation} 
where $H^{(1)}$ is the Hankel function of first kind\footnote{Using  
the asymptotic behavior of the Hankel function 
\begin{equation} 
H^{(1)}_{\nu }(z)\rightarrow _{z\rightarrow +\infty } 
\sqrt{\frac{2}{\pi z}}{\rm e}^{i(z-\pi \nu /2-\pi /4)}\, , 
\end{equation} 
one can easily check that the mode function $\mu _{_{\rm S,T}}^{\rm 
stand}(k,\eta )$ has indeed the required limit given by 
Eq.~(\ref{stic}).}. This result allows us to calculate the power 
spectra. For density perturbations and gravitational waves, one 
respectively finds, on super-Hubble scales 
\begin{eqnarray} 
k^3P_{\zeta}(k) &=& \frac{\ell _{_{\rm Pl}}^2}{\ell _0^2} 
\frac{1}{\pi \gamma }f(q)k^{2q+2}\, ,  
\\ 
k^3P_{h}(k) &=& \frac{\ell _{_{\rm Pl}}^2}{\ell _0^2} 
\frac{16}{\pi }f(q)k^{2q+2}\, , 
\end{eqnarray} 
where $\ell _{_{\rm Pl}}=m_{_{\rm Pl}}^{-1}$ is the Planck length  
and the function $f(q)$ is given by 
\begin{equation} 
f(q)\equiv \frac{1}{\pi }\biggl[\frac{\Gamma (1/2-q)}{2^{q}}\biggr]^2\, . 
\end{equation} 
In the above definition of the function $f(q)$, $\Gamma $ denotes 
Euler's integral of the second kind. For the de Sitter case, one has 
$f(q=-1)=1$. However, for this case the amplitude of density 
perturbations blows up since $\gamma (q=-1)=0$ while the amplitude of 
the gravitational waves spectrum remains finite. The origin of the 
singular limit for density perturbations is again the factor 
$\sqrt{\gamma }$ at the denominator of the expression for the scalar 
power spectrum. In fact, this case must be analyzed separately and one 
can show that density perturbations do not exist in de Sitter 
space-time. We see that power-law inflation leads to 
\begin{equation} 
k^3P_h(k) = A_{_{\rm T}}k^{n_{_{\rm T}}} \,\, , \,\,  
k^3P_{\zeta }(k) = A_{_{\rm S}}k^{n_{_{\rm S}}-1} 
\end{equation}  
with $n_{_{\rm S}}-1=n_{_{\rm T}}$. Explicitly, 
one has 
\begin{equation} 
n_{_{\rm S}}=2q+3\, . 
\end{equation} 
The case $q=-1$ leads to a scale-invariant spectral index, namely  
$n_{_{\rm S}}=1$ and $n_{_{\rm T}}=0$. 
 
\par 
 
The power-spectrum of the scalar field can be also calculated and  
the result reads (contrary to the previous power spectra, the power  
spectrum of the scalar field is a dimension-full quantity) 
\begin{equation} 
k^3P_{\chi }(k)=\frac{1}{(2\pi )^2\ell _0^2}f(q)k^{2q+2}\, . 
\end{equation} 
It is also convenient to expressed this power spectrum in terms of the 
Hubble parameter during inflation $H_{\rm inf}={\cal H}/a= -q/[\ell 
_0(-\eta )^{q+1}]$. This permits to replace the scale $\ell _0$ in the 
above equation and leads to 
\begin{equation} 
k^3P_{\chi }(k)=\frac{H_{\rm inf}^2(\eta )}{(2\pi )^2} 
\frac{(-\eta )^{2(q+1)}}{q^2}f(q)k^{2q+2}\, . 
\end{equation} 
Of course, in spite of the fact that the time dependence now appears 
explicitly, the power spectrum remains a time independent quantity. 
In the de Sitter case, $q=-1$, one recovers the result often 
cited in the literature, namely $k^3P_{\chi }=[H_{\rm inf}/(2\pi 
)]^2$. The spectrum becomes scale-invariant and its amplitude remains 
finite. The situation, except for some unimportant numerical factors, 
is very similar to that of gravitational waves.  
 
To conclude this subsection, let us emphasize that the previous 
discussion has shown that 
the analogy between density perturbations and a scalar field must be 
used cautiously. The spectral indices are similar even in the de 
Sitter limit but the amplitudes differ radically in this limit. 
 
\subsection{Slow-roll inflation} 
 
The slow-roll method is an approximation scheme which allows us to 
go beyond the simple power-law solutions considered in the previous 
section. It permits to treat a more general class of inflaton 
potentials. At leading order, the approximation is controlled by the 
slow-roll parameters (see e.g.~Ref.~\cite{Lea}; for a new set of 
slow-roll parameters with a very nice interpretation, see 
Ref.~\cite{ST}) defined by: 
\begin{eqnarray} 
\label{defepsilon} 
\epsilon &\equiv & 3 \frac{\dot{\varphi_0}^2}{2}  
\left(\frac{\dot{\varphi _0}^2}{2} + V\right)^{-1} = -\frac{\dot{H}}{H^2} 
=1-\frac{{\cal H}'}{{\cal H}^2}\, , 
\nonumber \\ 
\label{defdelta} 
\delta &\equiv & -\frac{\ddot{\varphi }}{H\dot{\varphi }} = 
- \frac{\dot{\epsilon }}{2 H \epsilon }+\epsilon \, ,\quad  
\label{defxi} 
\xi \equiv \frac{\dot{\epsilon }-\dot{\delta }}{H}\ . 
\end{eqnarray} 
The quantity $V(\varphi )$ is the inflaton potential. We see that 
$\gamma =\epsilon$, where $\gamma $ is the function that has been 
introduced before. The slow-roll conditions are satisfied if 
$\epsilon$ and $\delta$ are much smaller than one and if $\xi = {\cal 
O}(\epsilon^2,\delta^2,\epsilon\delta)$. Since the equations of motion 
for $\epsilon$ and $\delta$ can be written as: 
\begin{equation} 
\label{eqmotionsrpara} 
\frac{\dot{\epsilon}}{H}=2\epsilon (\epsilon -\delta)\ , \quad  
\frac{\dot{\delta }}{H}=2\epsilon (\epsilon -\delta )-\xi \, , 
\end{equation} 
it is clear that this amounts to considering $\epsilon$ and $\delta$ as 
constants if one works at first order in the slow-roll 
parameters. This property turns out to be crucial for the calculation 
of the power spectra of cosmological perturbations. For power-law 
inflation the slow-roll parameters satisfy: $\epsilon = \delta < 1$, 
$\xi = 0$. 
 
\par 
 
The slow-roll approximation can be viewed as a kind of expansion 
around the de Sitter space-time. Indeed, at leading order, one has $aH 
\approx -(1+\epsilon)/\eta$ which implies that the scale factor 
behaves like $a(\eta)\propto (-\eta )^{-1-\epsilon}$.  Interestingly 
enough, the effective power index at leading order depends on 
$\epsilon$ only. At this point, one should make the following 
remark. Using the relation $\gamma =(q+1)/q$, the scale factor of 
power-law inflation can be re-written as $a(\eta )\propto (-\eta 
)^{-1/(1-\gamma )}$. From this expression, one might be tempted to 
write that, in the slow-roll framework, $a(\eta )\propto (-\eta 
)^{-1/(1-\epsilon )}$ since $\gamma =\epsilon $ in this 
approximation. Of course, this expression is inconsistent because it 
contains an infinite number of power of $\epsilon $. What should be 
done is to expand the expression $-1/(1-\epsilon )$ in terms of 
$\epsilon$ from which we recover that the scale factor is given by 
$a(\eta)\propto (-\eta )^{-1-\epsilon}$. As long as one decides to 
keep high order terms, the whole hierarchy of slow-roll parameters 
should enter the game. This shows that the slow-roll approximation 
does not only consist in naively expanding the power index of the 
scale factor in powers of $\epsilon $. This conclusion is reinforced by a study 
of the cosmological perturbations within this approximation. 
 
\par 
 
The effective potential of density perturbations can be calculated 
exactly in terms of the slow-roll parameters. The result is: 
\begin{equation} 
\label{potsr} 
U_{_{\rm S}}(\eta) = a^2H^2 [2 - \epsilon +  
(\epsilon -\delta)(3-\delta )+\xi] \ . 
\end{equation} 
We have seen before that, in the slow-roll approximation, $a^2 H^2 
\approx \eta^{-2}(1+ 2\epsilon)$. This implies that the effective 
potential reduces to 
\begin{equation} 
U_{_{\rm S}}(\eta )\approx \frac{1}{\eta ^2}(2+6\epsilon -3\delta )\, .  
\end{equation} 
Since, at leading order, $\epsilon$ and $\delta$ must be seen as 
constants in the slow-roll approximation, the equation of motion is of 
the same type as in power-law inflation and the solution is expressed 
in terms of Bessel functions according to: 
\begin{equation} 
\label{regionII} 
\mu _{_{\rm S}}=(k\eta )^{1/2}[B_1J_{q_{_{\rm S}}-1/2}(k\eta ) 
+B_2J_{-q_{_{\rm S}}+1/2}(k\eta )]. 
\end{equation} 
The parameter $q_{_{\rm S}}$ appearing in the order of the Bessel 
function is given by 
\begin{equation} 
\label{orderS} 
q_{_{\rm S}}= -1-2\epsilon +\delta \ .  
\end{equation} 
A comment is in order here: The potential $U_{_{\rm S}}$ depends on 
the scale factor and its derivatives only and we have seen before that 
the scale factor behaves as $a(\eta )\propto (-\eta )^{-1-\epsilon }$. 
Therefore, one might think that $U_{_{\rm S}}$ should depend on 
$\epsilon$ only. This is not the case. The reason is that $U_{_{\rm 
S}}$ contains terms like $\dot{\epsilon }/\epsilon $ (for instance) 
which are linear in $\delta$. First one must calculate all 
derivatives, replace them with their expression in terms of $\epsilon$ 
and $\delta$, and only then consider that the slow-roll parameters are 
constant. For gravitational waves, the same lines of reasoning can be 
applied. The effective potential can be written as $U_{_{\rm T}}(\eta 
) = a^2 H^2 \left(2 - \epsilon\right)$ and gives in the slow-roll 
limit 
\begin{equation} 
\label{potgwsr} 
U_{_{\rm T}}(\eta ) \sim \frac{2 + 3\epsilon}{\eta ^2} \ . 
\end{equation} 
Therefore, the solution of $\mu _{_{\rm T}}$ is similar to the one 
given in Eq.~(\ref{regionII}), where the effective index of the Bessel 
function is now given by: $q_{_{\rm T}}=-1-\epsilon $. This time, the 
spectral index only depends on $\epsilon $ as expected from the shape 
of the tensor effective potential. Fixing the initial conditions in 
the infinite past in the standard manner, we arrive at the following 
expressions for the power spectra 
\begin{widetext} 
\begin{eqnarray} 
\label{srdp} 
{\cal P}_{\zeta }(k) &=& \frac{H^2}{\pi\epsilon m_{_{\rm Pl}}^2} 
\left[1 - 2(C+1)\epsilon - 2C(\epsilon -\delta ) 
  - 2(2\epsilon -\delta )\ln\left(\frac{k}{k_*}\right)\right]\, ,  
\\ 
\label{srgw} 
{\cal P}_h(k) &=& \frac{16 H^2}{\pi m_{_{\rm Pl}}^2} 
\left[1 - 2(C+1)\epsilon - 2\epsilon \ln  
\left(\frac{k}{k_*}\right)\right]\, ,  
\end{eqnarray}  
\end{widetext} 
where $C\equiv \gamma _{_{\rm E}}+\ln 2-2\simeq - 0.7296$, $\gamma
_{_{\rm E}}\simeq 0.5772$ being the Euler constant. All quantities are
evaluated at Hubble radius crossing. The scale $k_*$ is called the
pivot scale, see Refs.~\cite{MS2} and \cite{MRS}. One sees again that
the de Sitter limit $\epsilon \rightarrow 0$ is ill-defined for scalar
metric fluctuations, whereas it is well-defined for gravitational
waves.
 
\section{Minimal trans-Planckian physics} 
 
\subsection{General description} 
 
\begin{figure*}[t] 
\includegraphics*[width=18cm, height=10cm, angle=0]{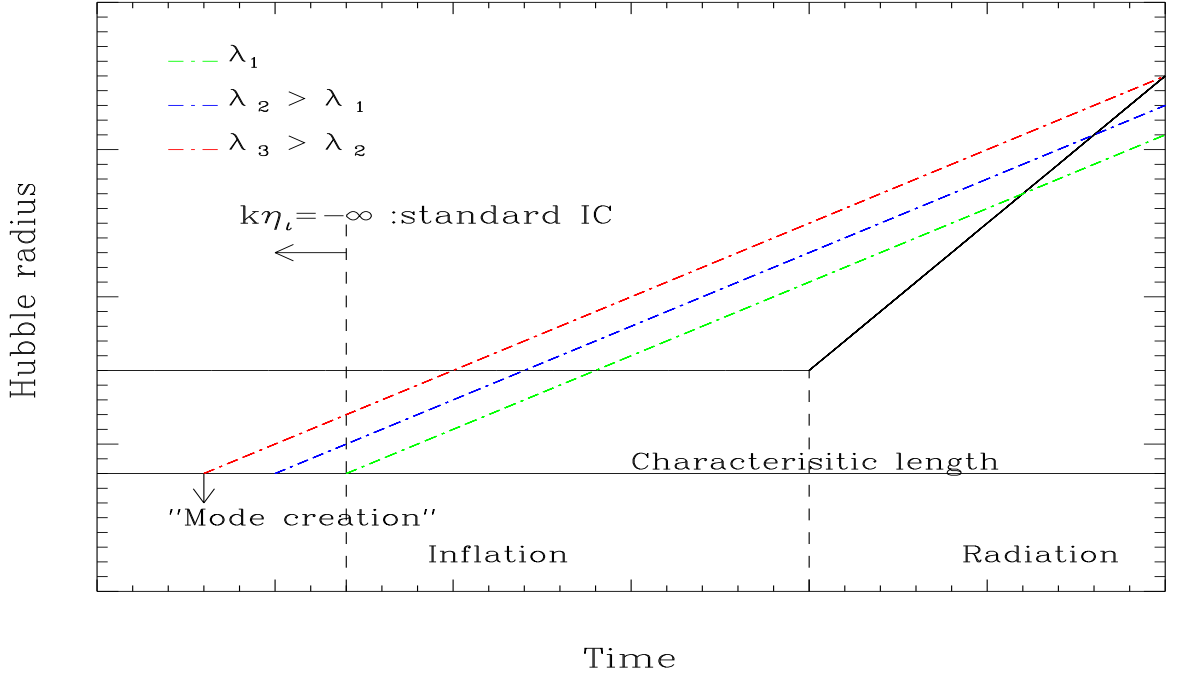} 
\caption{Sketch of the evolution of the Hubble radius vs time 
comparing how the initial conditions are fixed in the standard 
procedure and in the framework of the minimal trans-Planckian 
physics. In the standard procedure, the initial conditions are fixed 
on a surface of constant time $\eta =\eta _{\rm i}$ with $\eta _{\rm 
i}\rightarrow -\infty$. In this sense, the initial time does not 
depend on the wavenumber. On the contrary, in the minimal 
trans-Planckian physics, the initial time depends on the wavenumber 
and the Fourier modes are ``created'' when their wavelength is equal 
to a characteristic length. They never penetrate into the 
trans-Planckian region. This causes a modification of the power 
spectrum which is $k$ dependent since the modification is not the same 
for all Fourier modes as is apparent from the figure. } 
\label{tpl3} 
\end{figure*}

We now consider the cosmological perturbations of quantum-mechanical 
origin in the framework of the minimal trans-Planckian physics. The 
new ingredient consists in assuming that the Fourier modes never 
penetrate the trans-Planckian region. Rather, the main idea is that a 
Fourier mode is ``created'' when its wavelength becomes equal to a new 
fundamental characteristic scale $\ell _{_{\rm C}}\equiv (2\pi 
)/M_{_{\rm C}}$. Then, the evolution proceeds as usual since the 
equations of motion of the mode functions $\mu _{_{\rm S,T}}$ are 
taken to be unmodified and still given by Eqs.~(\ref{eomtensor}) and 
(\ref{eomscalar}). In this way, the changes are entirely encoded 
in the initial conditions and there is no need to postulate some 
ad-hoc trans-Planckian physics. The time of ``creation'' of the mode 
of comoving wavenumber $k$, $\eta _k$, can be computed from the 
condition 
\begin{equation} 
\lambda (\eta _k)=\frac{2\pi }{k}a(\eta _k)= 
\frac{2\pi }{M_{_{\rm C}}}=\ell _{_{\rm C}}\, , 
\end{equation} 
which implies that $\eta _k$ is a function of $k$. This has to be 
compared with the standard calculations where the initial time is 
taken to be $\eta _k=-\infty $ for any Fourier mode $k$ (in a  
certain sense, this initial time does not depend on $k$). The  
situation is summarized in Fig.~\ref{tpl3} 
 
\par 
 
In the framework described above, a crucial question is in which state 
the Fourier mode is created at the time $\eta _k$. There is now no 
asymptotic region in the infinite past where the standard 
prescriptions can be applied. In this article, we consider the most 
general conditions, namely that the mode is placed in an $\alpha 
$-vacuum according to  
\begin{eqnarray} 
\label{ci1} 
\mu _{_{\rm S,T}}(\eta _k) &=& \mp 
\frac{c_k+d_k}{\sqrt{2\omega _{_{\rm S,T}}(\eta _k)}} 
\frac{4\sqrt{\pi }}{m_{_{\rm Pl}}}\, ,  
\\ 
\label{ci2} 
\mu '_{_{\rm S,T}}(\eta _k) &=& 
\pm i\sqrt{\frac{\omega _{_{\rm S,T}}(\eta _k)}{2}} 
\frac{4\sqrt{\pi }(c_k-d_k)}{m_{_{\rm Pl}}}\, , 
\end{eqnarray} 
where $\omega _{_{\rm S,T}}\equiv \sqrt{k^2-U_{_{\rm S,T}}}$ is the 
effective frequency for density perturbations and gravitational waves. 
The coefficients $c_k$ and $d_k$ are {\it a priori} two arbitrary 
complex numbers satisfying the condition $\vert c_k\vert ^2-\vert 
d_k\vert ^2=1$. The instantaneous Minkowski state corresponds to $c_k=1$ 
and $d_k=0$. If, in addition, we take $\eta _k=-\infty $, we recover 
the standard choice, see Eq.~(\ref{stic}) since $\omega _{_{\rm 
S,T}}(\eta _k=-\infty )=k$. We are now in a position where we can 
compute the mode functions and the corresponding power spectra for 
density perturbations and gravitational waves. 
 
\par 
 
\subsection{Calculation of the mode function} 
 
In this section, we write the scale factor as $a(\eta )\propto (-\eta 
)^{p}$, where $p$ is a generalized index defined by  
\begin{equation} 
p= 
\cases{ 
\displaystyle{q=-\frac{1}{1-\gamma }}\, ,& 
\mbox{power-law inflation}\, , 
\cr 
 & \cr 
-1-\epsilon\, , & \mbox{slow-roll inflation} \, . 
} 
\end{equation} 
Since the initial conditions given by Eqs.~(\ref{ci1}) and (\ref{ci2}) 
are different from the standard ones, we will obtain different mode 
functions. The new mode functions that we want to calculate can be 
expanded according to 
\begin{equation} 
\label{nonstmodef} 
\mu _{_{\rm S,T}}(\eta )=\alpha _{_{\rm S,T}}(k) \mu _{_{\rm 
S,T}}^{\rm stand}(\eta ) +\beta _{_{\rm S,T}}(k) [\mu _{_{\rm 
S,T}}^{\rm stand}(\eta )]^*\, , 
\end{equation} 
where $\mu _{_{\rm S,T}}^{\rm stand}(\eta )$ denotes the mode 
functions obtained in the standard situation and given by the 
expression 
\begin{equation} 
\mu _{_{\rm S,T}}^{\rm stand}(k,\eta )=\mp \frac{2i\pi }{m_{_{\rm 
Pl}}}(-\eta )^{1/2} {\rm e}^{-i\pi \nu/2}H^{(1)}_{1/2-\nu}(-k\eta )\, 
. 
\end{equation} 
In this equation, we have introduced another generalized index $\nu $  
defined by 
\begin{equation} 
\nu= 
\cases{ 
\displaystyle{q=-\frac{1}{1-\gamma }}\, ,& 
\mbox{power-law inflation}\, , 
\cr 
 & \cr 
q_{_{\rm S}}=-1-2\epsilon+\delta \, , & \mbox{slow-roll inflation} \, , 
\cr 
 & \cr 
q_{_{\rm T}}=-1-\epsilon \, , & \mbox{slow-roll inflation} \, . 
} 
\end{equation} 
For power-law inflation the generalized index is the same for density 
perturbations and for gravitational waves (hence we do not need to 
distinguish them in the above definition). For slow-roll inflation, as 
already mentioned above, they differ. 
 
\par 
 
The coefficients $\alpha _{_{\rm S,T}}(k)$ and $\beta _{_{\rm 
S,T}}(k)$ are readily obtained using the initial conditions given in 
Eqs.~(\ref{ci1}) and (\ref{ci2}) 
\begin{widetext} 
\begin{eqnarray} 
\label{alpha} 
\alpha _{_{\rm S,T}}(k) &=& \frac{1}{4}(c_k+d_k) 
{\rm e}^{i\pi \nu/2} 
\sqrt{\frac{\pi }{-2\omega _{_{\rm S,T}}(\eta _k)\eta _k}} 
\biggl\{k\eta_k\biggl[H^{(2)}_{3/2-\nu}-H^{(2)}_{-1/2-\nu} 
\biggr] 
+\biggl[1+2i\frac{c_k-d_k}{c_k+d_k}\omega _{_{\rm S,T}}(\eta _k)\eta _k 
\biggr]H^{(2)}_{1/2-\nu}\biggr\}\, , 
\\ 
\label{beta} 
\beta _{_{\rm S,T}} (k) &=& \frac{1}{4}(c_k+d_k) 
{\rm e}^{-i\pi \nu/2} 
\sqrt{\frac{\pi }{-2\omega _{_{\rm S,T}}(\eta _k)\eta _k}} 
\biggl\{k\eta _k\biggl[H^{(1)}_{3/2-\nu}-H^{(1)}_{-1/2-\nu}\biggr] 
+\biggl[1+2i\frac{c_k-d_k}{c_k+d_k}\omega _{_{\rm S,T}}(\eta _k)\eta _k 
\biggr]H^{(1)}_{1/2-\nu}\biggr\}\, , 
\end{eqnarray} 
\end{widetext} 
where the Hankel functions are evaluated at $-k\eta _k$. The knowledge 
of the coefficients $\alpha _{_{\rm S,T}}(k)$ and $\beta _{_{\rm 
S,T}}(k)$ is equivalent to the knowledge of the modified mode 
function, see Eq.~(\ref{nonstmodef}). The quantity $-k\eta _k$ can be 
written as $-k\eta _k=-p/\sigma _k$ where $\sigma _k \equiv H(\eta 
_k)/M_{_{\rm C}}$. The explicit expression of $\sigma _k$ can be 
easily derived. Writing that $H=p/(a\eta )$, we find that $\sigma _k 
=\sigma _0(\eta _k/\eta _0)^{-p-1}$ where the time $\eta _0$ is a 
given, {\it a priori} arbitrary, time during inflation which, and this 
is the important point, does not depend on $k$. The quantity $\sigma 
_0$ is defined by $\sigma _0\equiv H_0/M_{_{\rm C}}$. Using that $\eta 
_k=p/(k\sigma _k)$ and $\eta _0=p/(\sigma _0a_0M_{_{\rm C}})$ we 
finally arrive at 
\begin{equation} 
\sigma _k=\sigma _0\biggl(\frac{k}{a_0M_{_{\rm C}}}\biggr)^{-1-1/p}\, . 
\end{equation} 
In the case of slow-roll inflation, the previous expression reduces to 
$\sigma _k=\sigma _0[k/(a_0M_{_{\rm C}})]^{-\epsilon }$. For the de 
Sitter case, $p=-1$ and $\epsilon =0$, the Hubble parameter and 
therefore $\sigma _k$ are constant. This means that $\eta _k\propto 
1/k$ as expected. An important remark is that $\sigma _k$ only depends 
on $p$. This means that, in the slow-roll approximation, $\sigma _k$ 
only depends on the slow-roll parameter $\epsilon $ and not on the 
slow-roll parameter $\delta $. This is due to the fact that the 
calculation of $\sigma _k$ only depends on background quantities. The 
slow-roll parameter $\delta $ appears in the calculation through the 
generalized index $\nu $, i.e. at the perturbed level. We now discuss 
the spectrum obtained according to the initial state chosen. 
 
\subsection{Instantaneous Minkowski vacuum} 
 
\begin{figure*}[t] 
\includegraphics*[width=18cm, height=10cm, angle=0]{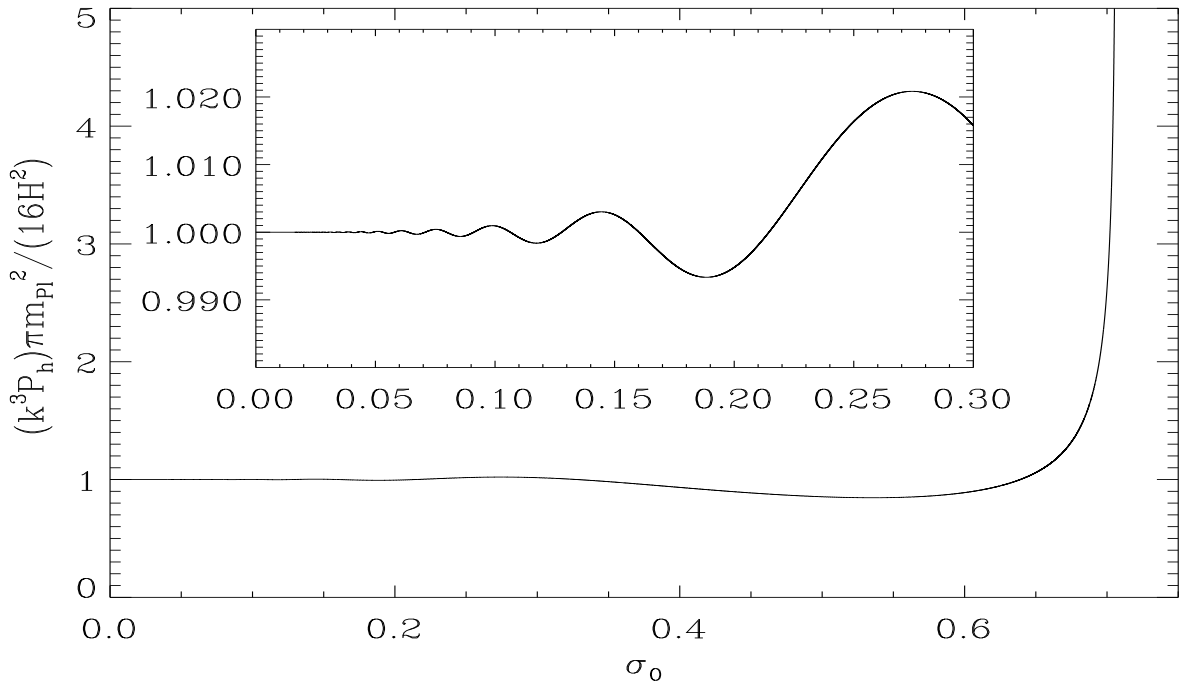} 
\caption{Amplitude of the gravitational wave power spectrum as a 
function of the parameter $\sigma _0$ in the case where the initial  
state is taken to be 
the instantaneous Minkowski state. For small values of $\sigma _0$, the 
leading order correction is cubic in $\sigma _0$. The amplitude blows  
up when $\sigma _0=1/\sqrt{2}$.} 
\label{tpl1} 
\end{figure*}

In this section, we assume that the initial state is such that $c_k=1$ 
and $d_k=0$. As a warm up, let us derive the spectrum in the 
particular case of de Sitter, $p=-1$. As mentioned above, this can 
only been done for gravitational waves since the de Sitter limit is 
singular for density perturbations. For $p=-1$, we have $\sigma 
_k=\sigma _0$ and $-k\eta _k=1/\sigma _0$. The spectrum can be 
determined exactly because the Bessel functions in Eqs.~(\ref{alpha}) 
and (\ref{beta}) reduce to ordinary functions in the case $p=-1$. The 
result reads 
\begin{widetext} 
\begin{eqnarray} 
\label{alTdS} 
\alpha _{_{\rm T}}(k) &=& \frac{1}{2}{\rm e}^{-i\pi /2-i/\sigma _0} 
(1-2\sigma _0^2)^{-1/4}\biggl[i+\sigma _0-i\sigma _0^2+(i+\sigma _0) 
\sqrt{1-2\sigma _0^2}\biggr]\, , 
\\ 
\label{beTdS} 
\beta _{_{\rm T}}(k) &=& \frac{1}{2}{\rm e}^{i\pi /2+i/\sigma _0} 
(1-2\sigma _0^2)^{-1/4}\biggl[-i+\sigma _0+i\sigma _0^2+(i-\sigma _0) 
\sqrt{1-2\sigma _0^2}\biggr]\, . 
\end{eqnarray} 
The modulus of the two previous expressions can be easily  
derived from the above relation 
\begin{equation} 
\vert \alpha _{_{\rm T}}\vert = \frac{1}{2} 
\biggl(\frac{2-2\sigma _0^2-\sigma _0^4} 
{\sqrt{1-2\sigma _0^2}}+2\biggl)^{1/2}\, , \quad  
\vert \beta _{_{\rm T}}\vert = \frac{1}{2} 
\biggl(\frac{2-2\sigma _0^2-\sigma _0^4} 
{\sqrt{1-2\sigma _0^2}}-2\biggl)^{1/2}\, , 
\end{equation} 
which coincides with the result of Ref.~\cite{npc}. One can check that 
$\vert \alpha _{_{\rm T}}\vert ^2-\vert \beta _{_{\rm T}}\vert 
^2=1$. In addition, one notices that the result is independent on $k$ 
as expected for the de Sitter space-time. Then, the power spectrum on 
super-horizon scales is given by $k^3P_h(k)=(2k^3/\pi ^2) \vert (\alpha 
_{_{\rm T}}\mu _{_{\rm T}}^{\rm stand}+\beta _{_{\rm T}}\mu _{_{\rm 
T}}^{\rm stand *} )/a\vert ^2$ where the super-horizon standard mode 
function can be expressed as $\mu _{_{\rm T}}^{\rm stand} \simeq 
-4i\sqrt{\pi }(2k)^{-1/2}/(m_{_{\rm Pl}}k\eta )$. The result reads 
\begin{equation} 
\label{minkops} 
k^3P_h(k)=\frac{16H^2_{\rm inf}}{\pi m_{_{\rm Pl}}^2} 
\frac{1}{\sqrt{1-2\sigma _0^2}}\biggl[1-\sigma _0^2-\frac{1}{2}\sigma _0^4 
+\sigma _0^3\sin \biggl(\frac{2}{\sigma _0}\biggr)+\frac{1}{2}\sigma _0^4 
\cos \biggl(\frac{2}{\sigma _0}\biggr)\biggr]\, . 
\end{equation} 
\end{widetext} 
This spectrum is represented in Fig.~\ref{tpl1}. 
So far, no approximation has been made. If we now assume that $\sigma 
_0$ is a small quantity, we can expand the result in terms of this 
quantity. Let us also remark that the terms $\sin (2/\sigma _0)$ and 
$\cos (2/\sigma _0)$ are non-analytic for small values of $\sigma 
_0$. At leading order, we obtain that $\vert \alpha _{_{\rm 
T}}\vert \sim 1$ and $\vert \beta _{_{\rm T}}\vert \sim \sigma _0^3/2$ 
in agreement with Ref.~\cite{npc}. This leads to the following 
spectrum 
\begin{equation} 
k^3P_h(k)\simeq \frac{16H_{\rm inf}^2}{\pi m_{_{\rm Pl}}^2} 
\biggl[1+\sigma _0^3\sin \biggl(\frac{2}{\sigma _0}\biggr)  
\biggr]\, . 
\end{equation} 
Therefore, the corresponding effect in the power spectrum is of order  
$\sigma _0^3$, i.e. a tiny effect if $\sigma _0$ is small. Note, however, 
that it is not possible to distinguish this effect from a change in 
the parameter $H_{\rm inf}$ of the inflationary background. 

\par 
 
We now turn to the case of slow-roll inflation. We start with density 
perturbations. The first step consists in calculating the coefficients 
$\alpha _{_{\rm S}}$ and $\beta _{_{\rm S}}$. At first order in the 
slow-roll parameters and at leading order in the parameter $\sigma _0$ 
we have 
\begin{eqnarray} 
\alpha _{_{\rm S}} &\simeq & {\rm e}^{ik\eta _k}\, ,  
\\ 
\beta _{_{\rm S}} &\simeq & i\biggl(\frac12  
-\frac34 \delta -\frac32 \epsilon \ln \frac{k}{a_0M_{_{\rm C}}}  
\biggr)\sigma _0^3{\rm e}^{-ik\eta _k}\, . 
\end{eqnarray}
It is interesting to calculate the modulus of the coefficient $\beta 
_{_{\rm S}}$. Straightforward calculations lead to 
\begin{equation} 
\label{modbeta}
\vert \beta _{_{\rm S}} \vert \simeq \biggl(\frac12  
-\frac34 \delta -\frac32 \epsilon \ln \frac{k}{a_0M_{_{\rm C}}}  
\biggr)\sigma _0^3 \, . 
\end{equation} 
We see that this quantity depends on both slow-roll parameters
$\epsilon $ and $\delta $, contrary to what was obtained in
Ref.~\cite{npc} where the coefficient $\beta _{_{\rm S}}$ was found
not to depend on the slow-roll parameter $\delta $. The main reason
for this discrepancy is that, in Ref.~\cite{npc}, the spectrum of a
scalar field on an unperturbed background cosmological model (or,
equivalently, the spectrum of gravitational waves) was calculated, and
not the spectrum of scalar metric fluctuations, see the discussion
after Eq.~(\ref{specscaf}). If one uses the expression of the scale
factor $a(\eta )\propto (-\eta )^{-1-\epsilon}$ and inserts this into
the expression for the power spectrum for a scalar field on an
unperturbed background, one does not obtain the correct expression for
the power spectrum of density perturbations and one misses the terms
proportional to the slow-roll parameter $\delta $. One only obtains
either the power spectrum of gravitational waves or the power spectrum
of density perturbations in the very particular case of power-law
inflation for which we have $\epsilon =\delta $ (and for which, in
fact, the exact solution is known). This is why putting $\epsilon
=\delta $ in Eq.~(\ref{modbeta}) reproduces the result found in
Ref.~\cite{npc}.
 
\par 
 
Then, the power spectrum of the conserved quantity $\zeta $ is given  
by the following expression [to be compared with (\ref{srdp})]
\begin{widetext} 
\begin{eqnarray} \label{eq54} 
k^3P_{\zeta } &=&\frac{H^2}{\pi \epsilon m_{_{\rm Pl}}^2} 
\Biggl(1-2(C+1)\epsilon -2C(\epsilon -\delta )-2(2\epsilon -\delta ) 
\ln \frac{k}{k_*}+\sigma _0^3\biggl\{\sin \biggl[\frac{2}{\sigma _0} 
\biggl(1+\epsilon +\epsilon \ln \frac{k}{a_0M_{_{\rm C}}}\biggr)\biggr] 
\nonumber \\ 
& & -\biggl[\frac32 \delta +3\epsilon \ln \frac{k}{a_0M_{_{\rm C}}} 
+2(C+1)\epsilon +2C(\epsilon -\delta )+2(2\epsilon -\delta ) 
\ln \frac{k}{k_*}\biggr]\sin \biggl[\frac{2}{\sigma _0} 
\biggl(1+\epsilon +\epsilon \ln \frac{k}{a_0M_{_{\rm C}}}\biggr)\biggr] 
\nonumber \\ 
& & -\pi (2\epsilon -\delta )\cos \biggl[\frac{2}{\sigma _0} 
\biggl(1+\epsilon +\epsilon \ln \frac{k}{a_0M_{_{\rm C}}}\biggr) 
\biggr]\biggr\} 
\Biggr)\, . 
\end{eqnarray} 
\end{widetext} 
Thus, the power spectrum has oscillations superimposed on the 
scale-invariant base. Hence, in this case the change in the 
spectrum is ``in principle'' physically measurable. However, 
the prospects for actually detecting the difference of order  
$\sigma_o^3$ between (\ref{eq54}) and (\ref{srdp}) are hopeless if the 
parameter $\sigma _0$ is small. Indeed if, for instance, we assume 
that $\sigma _0=10^{-2}$, a value consistent with string theory, and 
if we consider that $\epsilon \simeq \delta \simeq 10^{-1}$ then the 
correction to the power spectrum is of order $\simeq 10^{-7}$. There 
is no possibility to detect such a small effect even with the Planck 
satellite. 
 
\par 
 
For completeness, let us mention the slow-roll result for gravitational 
waves. In this case the coefficients $\alpha _{_{\rm T}}$ and $\beta 
_{_{\rm T}}$ can be deduced from $\alpha _{_{\rm S}}$ and $\beta 
_{_{\rm S}}$ by putting $\epsilon =\delta $. Then, the power spectrum  
is given by 
\begin{widetext} 
\begin{eqnarray} 
k^3P_h &=&\frac{16H^2}{\pi m_{_{\rm Pl}}^2} 
\Biggl(1-2(C+1)\epsilon -2\epsilon \ln \frac{k}{k_*} 
+\sigma _0^3\biggl\{\sin \biggl[\frac{2}{\sigma _0} 
\biggl(1+\epsilon +\epsilon \ln \frac{k}{a_0M_{_{\rm C}}}\biggr)\biggr] 
-\biggl[\frac32 \epsilon +3\epsilon \ln \frac{k}{a_0M_{_{\rm C}}} 
+2(C+1)\epsilon  
\nonumber \\ 
& & +2\epsilon  
\ln \frac{k}{k_*}\biggr]\sin \biggl[\frac{2}{\sigma _0} 
\biggl(1+\epsilon +\epsilon \ln \frac{k}{a_0M_{_{\rm C}}}\biggr)\biggr] 
-\pi\epsilon \cos \biggl[\frac{2}{\sigma _0} 
\biggl(1+\epsilon +\epsilon \ln \frac{k}{a_0M_{_{\rm C}}}\biggr) 
\biggr]\biggr\} 
\Biggr)\, . 
\end{eqnarray} 
\end{widetext} 
We can also estimate how the consistency check of inflation  
is modified. At first order in the slow-roll parameters and  
at leading order in the parameter $\sigma _0$, this quantity  
is the same as in the standard case, namely 
\begin{equation} 
R\equiv \frac{k^3P_h}{k^3P_{\zeta }}=16\epsilon \, . 
\end{equation} 
This is because at zeroth order in the slow-roll parameter, the  
leading contribution in $\sigma _0$ is the same for gravitational  
waves and density perturbations. 
 
\subsection{Arbitrary $\alpha$ vacuum state} 
 
We have established that, in the case of the instantaneous Minkowski
state, the correction to the power spectra is of order $\sigma
_0^3$. It is interesting to see whether this conclusion remains true
for other initial states. For this reason, we now repeat the
calculation of the Bogoliubov coefficients in the case of a de Sitter
space-time with an arbitrary $\alpha $ -vacuum state, characterized by
a value of $c_k$ and $d_k$, as the initial state. One finds
\begin{widetext} 
\begin{eqnarray} 
\alpha _{_{\rm T}}(k) &=& \frac{1}{2}(c_k+d_k){\rm e}^{-i\pi/2 -i/\sigma _0} 
(1-2\sigma _0^2)^{-1/4}\biggl[i+\sigma _0-i\sigma _0^2+z_k(i+\sigma _0) 
\sqrt{1-2\sigma _0^2}\biggr] \, ,  
\\ 
\beta _{_{\rm T}}(k) &=& \frac{1}{2}(c_k+d_k){\rm e}^{i\pi/2 +i/\sigma _0} 
(1-2\sigma _0^2)^{-1/4}\biggl[-i+\sigma _0+i\sigma _0^2+z_k(i-\sigma _0) 
\sqrt{1-2\sigma _0^2}\biggr] \, ,  
\end{eqnarray} 
where we have introduced the quantity $z_k$ defined by $z_k\equiv 
(c_k-d_k)/(c_k+d_k)$. On can easily check that for $c_k=1$ and 
$d_k=0$, i.e. $z_k=1$, the above expressions reduce to 
Eqs.~(\ref{alTdS}) and (\ref{beTdS}). One can also calculate the 
modulus of these Bogoliubov coefficients, 
\begin{eqnarray} 
\vert \alpha _{_{\rm T}}\vert ^2 &=& \frac{\vert c_k+d_k\vert ^2}{4} 
\biggl\{\frac{(1+z_kz_k^*)-\sigma _0^2(1+z_kz_k^*) 
+\sigma _0^4(1-2z_kz_k^*)}{\sqrt{1-2\sigma _0^2}} 
+[z_k^*+z-i(z_k^*-z_k)\sigma _0^3]\biggr\}\, ,  
\\ 
\vert \beta _{_{\rm T}}\vert ^2 &=& \frac{\vert c_k+d_k\vert ^2}{4} 
\biggl\{\frac{(1+z_kz_k^*)-\sigma _0^2(1+z_kz_k^*) 
+\sigma _0^4(1-2z_kz_k^*)}{\sqrt{1-2\sigma _0^2}} 
-[z_k^*+z_k+i(z_k^*-z_k)\sigma _0^3]\biggr\}\, ,  
\end{eqnarray} 
and verify that $\vert \alpha _{_{\rm T}} \vert ^2 - \vert \beta 
_{_{\rm T}} \vert ^2=\vert c_k\vert ^2-\vert d_k\vert ^2=1$. From the 
two previous relations, one can determine the power spectrum 
exactly. The result reads 
\begin{eqnarray} 
k^3P_h &=& \frac{16H^2_{\rm inf}}{\pi m_{_{\rm Pl}}^2} 
\frac{\vert c_k+d_k\vert ^2}{\sqrt{1-2\sigma _0^2}} 
\biggl\{\frac12 (1+z_kz_k^*)(1-\sigma _0^2) 
+\frac{\sigma _0^3}{2}(1-2z_kz_k^*)\biggl[\sigma _0 
-\sigma _0\cos \biggl(\frac{2}{\sigma _0}\biggr) 
-2\sin  \biggl(\frac{2}{\sigma _0}\biggr)\biggr] 
\nonumber \\ 
& & +\frac{1}{2}(1-z_kz_k^*)(3\sigma _0^2-1) 
\cos \biggl(\frac{2}{\sigma _0}\biggr) 
+(1-z_kz_k^*)\sigma _0\sin \biggl(\frac{2}{\sigma _0}\biggr) 
-\frac{i}{2}(z_k^*-z_k)(1-2\sigma _0^2)^{3/2} 
\sin \biggl(\frac{2}{\sigma _0}\biggr) 
\nonumber \\ 
& & -\frac{i}{2}\sigma _0^3(z_k^*-z_k)\sqrt{1-2\sigma _0^2} 
+\frac{i}{2}\sigma _0(z_k^*-z_k)(\sigma _0^2-2) 
\sqrt{1-2\sigma _0^2} 
\cos \biggl(\frac{2}{\sigma _0}\biggr)\biggr\}\, . 
\end{eqnarray} 
As is apparent from the above expression, the correction in the power
spectrum compared to the usual results is of order unity, i.e. not
suppressed by any power of $\sigma_0$. If one argues that the
Bogoliubov coefficients themselves should be of linear order in
$\sigma_0$ then the correction terms would also be linear in
$\sigma_0$. On the other hand, unless the Bogoliubov coefficients
depend on $k$, the effect is simply a change in the amplitude of the
spectrum, and can hence be absorbed in a redefinition of the
background. Thus, the effect is not physically measurable. On the
other hand, if the coefficients depend on $k$ (as discussed e.g. in
the analysis of \cite{Lowe1}) there is a large measurable effect on
the power spectrum.  For the instantaneous Minkowski state, one check
easily that only the two first terms in the curly brackets survives
and that the corresponding expression reduces to the one found
previously. If we expand this spectrum in powers of $\sigma_o$, we
find that
\begin{equation} 
k^3P_h = \frac{16H^2_{\rm inf}}{\pi m_{_{\rm Pl}}^2} 
\vert c_k+d_k\vert ^2\biggl\{ 
\biggl \vert z_k\cos \biggl(\frac{2}{\sigma _0}\biggr) 
-i\sin \biggl(\frac{2}{\sigma _0}\biggr)\biggr\vert ^2 
+\biggl[(1-z_kz_k^*)\sin \biggl(\frac{2}{\sigma _0}\biggr) 
-i(z_k^*-z_k)\cos \biggl(\frac{2}{\sigma _0}\biggr)\biggr]\sigma _0 
+\cdots \biggr\}\, . 
\end{equation} 
\end{widetext} 
We conclude that the result obtained in the previous subsection 
for the instantaneous Minkowske vacuum, 
namely that the 
corrections to the spectrum are suppressed by three powers of 
$\sigma_0$ appears to be very particular to the choice of that state.

\par 
 
However, the previous analysis does not cover all the possible 
cases. Indeed, as we are going to demonstrate, the proposal put 
forward in Ref.~\cite{Dan1} corresponds in fact to a case where the 
complex number $z_k$ can also depend on $\sigma _0$. The previous 
study assumed that $c_k$ and $d_k$ were pure complex 
numbers. Therefore, in Danielsson's case \cite{Dan1}, the analysis 
needs to be redone. 
 
\subsection{Danielsson's $\alpha $-vacuum state} 
 
We start our analysis with the simplest case, namely gravitational 
waves in a de Sitter background. We treat this case in some detail 
since then these calculations can be used to study more complicated 
situations, for instance slow-roll inflation. The Einstein-Hilbert 
action 
\begin{equation} 
S_{_{\rm E-H}}=(16\pi G)^{-1}\int R\sqrt{-g}{\rm d}^4x 
\end{equation} 
expanded to second order (since we are dealing with first order 
equations of motion) reads 
\begin{eqnarray} 
S_2 &=& -\frac{1}{16\pi G}\int \frac{1}{4}g^{\mu \nu} 
\partial _{\mu }(h^i{}_j)\partial _{\nu }(h^j{}_i) 
\sqrt{-g}\, {\rm d}^4x 
\\ 
&=&\frac{m_{_{\rm Pl}}^2}{64 \pi}\int \biggl[(h^i{}_j)'(h^j{}_i)'  
-\partial _k(h^i{}_j) \partial 
^k(h^j{}_i)\biggr]a^2(\eta ){\rm d}^4x\, . \nonumber 
\end{eqnarray} 
In the first of the two previous expressions, $g^{\mu \nu }$ is the
inverse of the FLRW metric. The next step consists in inserting the
expression for $h_{ij}$, see Eq.~(\ref{eq2}), into the action. This
gives
\begin{equation} 
S_2=-\frac{m_{_{\rm Pl}}^2}{16\pi}\sum _{s=1}^2 
\int {\rm d}^4x\biggl(\frac{1}{2} 
g^{\mu \nu}\partial _{\mu }h^s\partial _{\nu }h^s\biggr)\, , 
\end{equation} 
where the fields $h^s(\eta ,x^k)$, with $s = 1,2$, are defined by 
the following expression
\begin{eqnarray} 
h^s(\eta ,x^k)\equiv  
\frac{1}{a(\eta )} \frac{1}{(2\pi 
)^{3/2}}\int {\rm d}{\bf k}\,   
\mu ^s_{_{\rm T}}(\eta ){\rm e}^{i{\bf k}\cdot {\bf x}}\, .
\\
\nonumber
 \end{eqnarray} 
Therefore, the action of gravitational waves is equivalent to the
action of two decoupled scalar fields.  Inserting the Fourier
decomposition of the two scalar fields into the expression for the
action, one arrives at
\begin{widetext}
\begin{eqnarray} 
S_2 &=&\frac{m_{_{\rm Pl}}^2}{16\pi}\int {\rm d}\eta \sum _{s=1}^2
\int _{\setR^{3+}}{\rm d}^3{\bf k}\biggl\{ 
(\mu ^s_{_{\rm T}})'{}^*(\mu ^s_{_{\rm T}})' 
-\frac{a'}{a}\biggl[ 
(\mu ^s_{_{\rm T}})'(\mu ^s_{_{\rm T}})^*
+(\mu ^s_{_{\rm T}})'{}^*\mu ^s_{_{\rm T}}\biggr] 
+\biggl(\frac{a'{}^2}{a^2}-k^2\biggr) 
\mu ^s_{_{\rm T}}(\mu ^s_{_{\rm T}})^*\biggr\}\, . 
\end{eqnarray} 
\end{widetext}
We can check that this Lagrangian leads to the correct equation of
motion: we have (we did not take into account the overall constant;
here the bar over ${\cal L}$ means that we are considering the
Lagrangian in the Fourier space)
\begin{eqnarray} 
\frac{\delta \bar{\cal L}}{\delta \mu _{_{\rm T}}^s} &=& 
-2{\cal H} (\mu ^s_{_{\rm T}})'{}^*+2({\cal H}^2-k^2)
(\mu ^s_{_{\rm T}})'{}^* \, ,
\\
\frac{\delta \bar{\cal L}}{(\delta \mu _{_{\rm T}}^s)'}
&=& 2(\mu ^s_{_{\rm 
T}})'{}^*-2{\cal H} (\mu ^s_{_{\rm T}})^*.  
\end{eqnarray} 
Therefore, the Euler-Lagrange equations given by the well-known 
expression
\begin{equation}  
\frac{{\rm d}}{{\rm d}\eta} 
\biggl[\frac{\delta \bar{\cal L}}{(\delta \mu 
_{_{\rm T}}^s)'}\biggr] 
-\frac{\delta \bar{\cal L}}{\delta \mu _{_{\rm T}}^s}=0\, ,
\end{equation}  
reproduce the correct equations of motion for the variables $(\mu
^s_{_{\rm T}})^*$. Let us notice that we can also vary the Lagrangian
with respect to $(\mu ^s_{_{\rm T}})^*$ to obtain the equation of
motion for $\mu ^s_{_{\rm T}}$.
 
\par 
 
\begin{figure*}[t] 
\includegraphics*[width=18cm, height=10cm, angle=0]{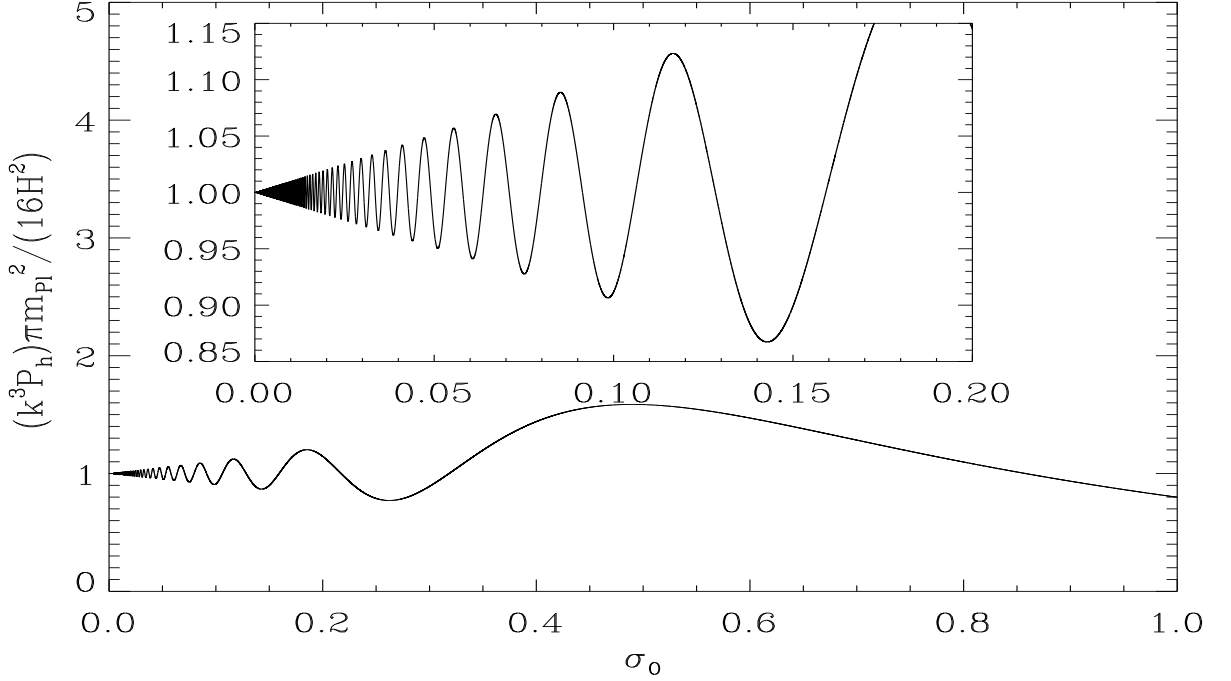} 
\caption{Amplitude of the gravitational wave power spectrum as  
a function of the parameter $\sigma _0$ in the case where the  
initial state is the one singled out by Danielsson's  
condition. For small values of $\sigma _0$, the leading order  
correction is linear in $\sigma _0$.} 
\label{tpl2} 
\end{figure*}

{}From the Lagrangian formalism that we have just described, we can 
now pass to the Hamiltonian formalism. The conjugate momentum to $\mu 
^s_{_{\rm T}}$ is defined by the formula 
\begin{equation} 
p^s_{_{\rm T}}=\frac{\delta \bar{{\cal L}}}{\delta (\mu ^s_{_{\rm T}})'{}^*} 
=\frac{m_{_{\rm Pl}}^2}{16\pi}\biggl[(\mu _{_{\rm T}}^s)'-\frac{a'}{a} 
\mu _{_{\rm T}}^s\biggl]\, . 
\end{equation} 
In real space, the conjugate momentum $\Pi ^s$ is defined  
by the expression  
\begin{equation} 
\Pi ^s(\eta ,x^k)=\frac{\delta {\cal L}}{\delta (h^s)'}=\frac{m_{_{\rm 
Pl}}^2}{16\pi} a^2(h^s)'\, . 
\end{equation} 
We can check that the two definitions are consistent by means of the 
relation 
\begin{equation} 
\Pi ^s(\eta ,x^k)=\frac{a(\eta )}{(2\pi )^{3/2}}\int  
{\rm d}{\bf k}p^s_{_{\rm T}}{\rm e}^{i{\bf k}\cdot {\bf x}}\, . 
\end{equation} 
Danielsson's boundary condition consists in demanding that, at the  
time of creation $\eta _k$, one has the usual relation characteristic  
of the standard vacuum state, namely 
\begin{equation} 
\Pi ^s(\eta _k,x^k)=-ik\frac{m_{_{\rm Pl}}^2}{16\pi } 
a^2(\eta _k)h^s(\eta _k,x^k)\, . 
\end{equation} 
Using the Fourier decomposition of the scalar fields $h^s$ and of 
their conjugate momenta, this last equation boils down to 
\begin{equation} 
(\mu _{_{\rm T}}^s)'-\frac{a'}{a}\mu _{_{\rm T}}^s=
-ik\mu _{_{\rm T}}^s\, . 
\end{equation} 
This relation is to be satisfied at the time $\eta =\eta _k$, and implies 
a link between the coefficients $\alpha _{_{\rm T}}$ and $\beta _{_{\rm 
T}}$. Since, in addition, $\vert \alpha _{_{\rm T}}\vert ^2-\vert 
\beta _{_{\rm T}}\vert ^2=1$, the coefficients are in fact completely 
fixed and they now depend on the parameter $\sigma _0$. Explicitly, 
the link can be expressed as 
\begin{equation} 
\label{ratioab} 
\frac{\beta _{_{\rm T}}}{\alpha _{_{\rm T}}}=- 
\displaystyle{\frac{(\mu _{_{\rm T}}^{\rm stand})' 
-(a'/a)\mu _{_{\rm T}}^{\rm stand}+ik \mu _{_{\rm T}}^{\rm stand}} 
{(\mu _{_{\rm T}}^{\rm stand*})' 
-(a'/a)\mu _{_{\rm T}}^{\rm stand*}+ik \mu _{_{\rm T}}^{\rm stand*}}}\, . 
\end{equation} 
So far no approximation has been made and the previous relation is 
general. We now restrict our study to the de Sitter case.  Using the 
explicit form of the mode function in this case, one arrives at 
\begin{eqnarray} 
\frac{\beta _{_{\rm T}}}{\alpha _{_{\rm T}}}= 
\frac{i}{i+2k\eta _k}{\rm e}^{-2ik\eta _k}\, , 
\\ \nonumber
\end{eqnarray} 
which is exactly the relation found in Ref.~\cite{Dan1}. From the 
normalization, we deduce that $\vert \alpha _{_{\rm T}}\vert 
^2=1+\sigma _0^2/4$. The exact power spectrum can now be determined 
since we explicitly know the coefficients $\alpha _{_{\rm T}}$ and 
$\beta _{_{\rm T}}$.  Performing the standard calculation, one finds 
\begin{equation} 
\label{dpsgw} 
k^3P_h =\frac{16H^2_{\rm inf}}{\pi m_{_{\rm Pl}}^2} 
\biggl[1+\frac{\sigma _0^2}{2}-\sigma _0 
\sin \biggl(\frac{2}{\sigma _0}\biggr) 
-\frac{\sigma _0^2}{2} 
\cos \biggl(\frac{2}{\sigma _0}\biggr)\biggr]\, . 
\end{equation} 
This expression should be compared with the corresponding equation 
found in the case of the Minkowski instantaneous state, see 
Eq.~(\ref{minkops}). In particular, expanding everything in terms of 
$\sigma _0$, one obtains 
\begin{equation} 
k^3P_h=\frac{16H^2_{\rm inf}}{\pi m_{_{\rm Pl}}^2} 
\biggl[1-\sigma _0 
\sin \biggl(\frac{2}{\sigma _0}\biggr)\biggr]\, , 
\end{equation} 
i.e., a first order effect instead of a third order effect. 
However, once again note that this effect can be reproduced 
by redefining the background cosmological parameters, and hence 
it is not a physically measurable effect. The  
spectrum is represented in Fig.~\ref{tpl2}. 
 
\par 
 
Let us now turn to the case of slow-roll inflation. The goal is simply 
to evaluate the ratio $\beta _{_{\rm T}}/\alpha _{_{\rm T}}$ given by 
Eq.~(\ref{ratioab}) to leading order in the parameter $\sigma _0$ and 
to first order in the slow-roll parameter $\epsilon$. One finds 
\begin{equation} 
\beta _{_{\rm T}}=-\frac{i\sigma _0}{2}{\rm e}^{-2ik\eta _k} 
\biggl(1-\epsilon \ln \frac{k}{a_0M_{_{\rm C}}}\biggr) 
\alpha _{_{\rm T}}\, . 
\end{equation} 
Of course, for $\epsilon =0$, one recover the exact result obtained 
previously at leading order in $\sigma _0$. Finally the power spectrum  
for gravitational waves reads 
\begin{widetext} 
\begin{eqnarray} \label{eq78} 
k^3P_h &=& \frac{16 H_{\rm inf}^2}{\pi m_{_{\rm Pl}}^2} 
\biggl\{1-2(C+1)\epsilon -2\epsilon \ln \frac{k}{k_*} 
-\sigma _0\biggl[1-2(C+1)\epsilon -2\epsilon \ln \frac{k}{k_*} 
-\epsilon \ln \frac{k}{a_0M_{_{\rm C}}}\biggr] 
\nonumber \\ 
& & \times \sin \biggl[\frac{2}{\sigma _0} 
\biggl(1+\epsilon +\epsilon \ln  
\frac{k}{a_0M_{_{\rm C}}}\biggr)\biggr]
+\sigma _0\pi \epsilon 
\cos \biggl[\frac{2}{\sigma _0} 
\biggl(1+\epsilon +\epsilon \ln  
\frac{k}{a_0M_{_{\rm C}}}\biggr)\biggr]\biggr\}\, . 
\end{eqnarray} 
\end{widetext} 
Note that since in this case the correction terms lead to 
oscillations in the spectrum about a scale-invariant 
base spectrum, the effect of the correction terms is 
in principle measurable. Since the amplitude is 
only suppressed by one power of $\sigma _0$, the 
prospects of being able to detect such effects in 
upcoming experiments are good.  
 
\par 
 
We now explore the same mechanism but for scalar metric (density)  
perturbations. The 
action is expressed in terms of the variable $\mu _{_{\rm S}}$ 
introduced before. It reads 
\begin{equation} 
S_2=\frac12 \int {\rm d}^4{\bf x}\biggl[(\mu _{_{\rm S}}')^2 
-\delta ^{ij}\partial _i\mu _{_{\rm S}}\partial _j\mu _{_{\rm S}} 
+\frac{(a\sqrt{\gamma })''}{a\sqrt{\gamma }}\mu _{_{\rm S}}^2\biggr]\, , 
\end{equation} 
In Fourier space, the action reads 
\begin{equation} 
S_2=\frac12 \int {\rm d}\eta \int _{\setR^+}{\rm d}^3{\bf k}\biggl\{ 
\mu _{_{{\rm S}}}'\mu _{_{{\rm S}}}^*{}'-\biggl[k^2- 
\frac{(a\sqrt{\gamma })''}{a\sqrt{\gamma }}\biggr] 
\mu _{_{{\rm S}}}\mu _{_{{\rm S}}}^*\biggr\}\, . 
\end{equation} 
The conjugate momentum to the variable $\mu _{_{\rm S}}$ is given by 
$\Pi _{\mu _{_{\rm S}}}=\mu _{_{\rm S}}'$ and therefore Danielsson's 
boundary condition reads 
\begin{equation} 
\label{condscalar} 
\mu _{_{{\rm S}}}'=-ik\mu _{_{{\rm S}}}\, ,  
\end{equation} 
where this relation must be evaluated at the time $\eta =\eta _k$.  We 
note that the term $(a'/a)\mu _{_{\rm S}}$, which was present in the 
case of the scalar field, does not appear for density 
perturbations. This can be traced back to the fact that the two 
quantities that are quantized, $\mu _{_{\rm T}}$ and $\mu _{_{\rm S}}$ 
do not have the same action. This means that the link between $\alpha 
_{_{\rm S}}$ and $\beta _{_{\rm S}}$ for density perturbations is not 
the same as for gravitational waves. For scalar metric 
fluctuations, it reads 
\begin{equation} 
\label{bascalar} 
\frac{\beta _{_{\rm S}}}{\alpha _{_{\rm S}}}=- 
\displaystyle{\frac{(\mu _{_{\rm S}}^{\rm stand})' 
+ik \mu _{_{\rm S}}^{\rm stand}} 
{(\mu _{_{\rm S}}^{\rm stand*})' 
+ik \mu _{_{\rm S}}^{\rm stand*}}}\, . 
\end{equation} 
This equation should be compared with Eq.~(\ref{ratioab}). The absence  
of the terms $a'/a$ has important consequences. In order to guess  
what the difference is, we can apply the previous equation to the  
de Sitter case, even if in principle this is not allowed (see the  
discussion above). One finds 
\begin{equation} 
\frac{\beta _k}{\alpha _k}=\frac{i}{i+2k\eta _k-2ik^2\eta ^2} 
{\rm e}^{-2ik\eta _k}\, . 
\end{equation} 
Expanding the previous expression in $\sigma _0$, we find that the 
first order terms cancel out and we are left with $\beta _k/\alpha _k 
\simeq -(\sigma _0^2/2){\rm e}^{2i/\sigma _0}$. Therefore, the result 
will be of order $\sigma _0^2$ and not of order $\sigma _0$ as it was 
the case for a scalar field and gravitational waves.  We would have 
obtained a linear correction in the power spectrum if Danielsson's 
condition had been 
\begin{equation} 
\biggl(\frac{\mu _{_{\rm S}}}{a}\biggr)'=-i\frac{k}{a}\mu _{_{\rm S}}\, , 
\end{equation} 
as it is for gravitational waves, instead of $\mu _{_{\rm S}}'=-ik\mu 
_{_{\rm S}}$, see Eq.~(\ref{condscalar}). 
 
\par 
 
\begin{figure*}[t] 
\includegraphics*[width=18cm, height=10cm, angle=0]{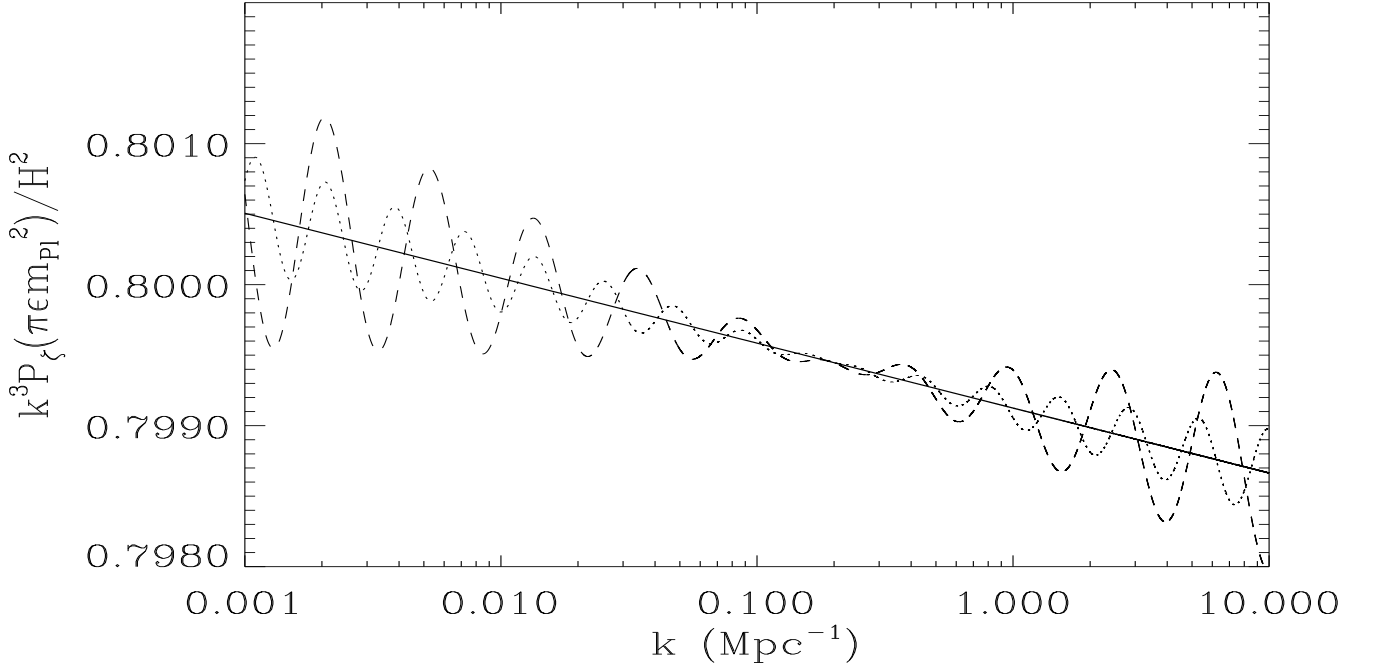} 
\caption{Power spectra given by Eq.~(\ref{pssrs2}). The slow-roll 
parameters are taken to be $\epsilon=0.10005$ and $\delta =0.2$ which 
corresponds to a tiny red tilt. The pivot scale is chosen to be 
$k_*=0.01 \mbox{Mpc}^{-1}$. The time $\eta _0$ is chosen to be the 
time at which the pivot scale is ``created''. This corresponds to 
$k_*/a_0=M_{_{\rm C}}$. The solid line corresponds to $\sigma _0=0$  
while the dotted line is for $\sigma _0=0.02$ and the dashed line  
for $\sigma _0=0.03$.} 
\label{tpl4} 
\end{figure*}

Let us now evaluate Eq.~(\ref{bascalar}) consistently and rigorously 
in the slow-roll approximation.  One finds 
\begin{equation} 
\beta _{_{\rm S}}=-\frac{\sigma _0^2}{2}{\rm e}^{-2ik\eta _k} 
\biggl(1+\epsilon -\frac32\delta -2\epsilon  
\ln \frac{k}{a_0M_{_{\rm C}}} \biggr)\alpha _{_{\rm S}}\, . 
\end{equation} 
As expected, the result is quadratic in $\sigma _0$. Repeating the 
standard calculations, one can find the explicit expression for the 
power spectrum 
\begin{widetext} 
\begin{eqnarray} 
\label{pssrs2} 
k^3P_{\zeta } &=&\frac{H^2}{\pi \epsilon m_{_{\rm Pl}}^2} 
\biggl\{1-2(C+1)\epsilon -2C(\epsilon -\delta )-2(2\epsilon -\delta ) 
\ln \frac{k}{k_*}+\sigma _0^2\biggl[ 
1-2(C+1)\epsilon -2C(\epsilon -\delta ) 
+\epsilon -\frac32\delta  
\nonumber \\ 
& & -2(2\epsilon -\delta )\ln \frac{k}{k_*} 
-2\epsilon \ln \frac{k}{a_0M_{_{\rm C}}} 
\biggr]\cos \biggl[\frac{2}{\sigma _0} 
\biggl(1+\epsilon +\epsilon \ln \frac{k}{a_0M_{_{\rm C}}}\biggr)\biggr] 
+\sigma _0^2\pi (2\epsilon -\delta )
\sin \biggl[\frac{2}{\sigma _0} 
\biggl(1+\epsilon +\epsilon \ln \frac{k}{a_0M_{_{\rm C}}}\biggr)\biggr] 
\biggr\} \, .
\nonumber \\ 
\end{eqnarray} 
\end{widetext} 
As in the case of gravitational waves, in principle this is a measurable 
effect since it corresponds to oscillations of the power spectrum.  
Typical power spectra are represented in Fig.~\ref{tpl4}. Since the  
correction is quadratic in $\sigma _0$, the correction turns out  
to be extremely small and the prospects for the detection of such  
an effect are not optimistic even with a high-accuracy experiment  
like Planck. 

Thus, at least with this choice of the precise form of the
action~\cite{Dprivate}, Danielsson's prescription leads to
trans-Planckian corrections of different strengths for gravitational
waves and for scalar fluctuations. Note that this also implies that
the consistency relation of inflation will be modified. Comparing
(\ref{eq78}) and (\ref{pssrs2}), at leading order in $\sigma _0$, one
finds
\begin{equation} 
R=16\epsilon \biggl\{1-\sigma _0\sin \biggl[\frac{2}{\sigma _0} 
\biggl(1+\epsilon +\epsilon \ln \frac{k_*}{a_0M_{_{\rm C}}} 
\biggr)\biggr]\biggr\}\, , 
\end{equation} 
where the ratio has been evaluated at the pivot scale. As expected 
from the previous considerations, the correction is linear in $\sigma 
_0$. 
 
\section{Comparison with analyses using modified dispersion relations} 
 
In this section, we consider a case where the trans-Planckian physics 
is described by mean of a modified dispersion relation. Although, this 
phenomenological description is different from the one considered 
above, it has already been shown that the corresponding spectrum can 
also possess superimposed oscillations \cite{MB1,BM2,LLMU}. In this 
section, we calculate the spectrum of a scalar field (or gravitational 
waves) living in a de Sitter space-time and we compare this result 
with the corresponding result obtained in the framework described in 
the previous sections, see Eqs.~(\ref{minkops}) and (\ref{dpsgw}). In 
particular, our goal is to determine which parameter controls the 
magnitude of the correction to the standard power spectrum in the case 
of a modified dispersion relation (for the case treated before, it was 
the parameter $\sigma _0$).  
 
\par 
 
\begin{figure*}[t] 
\includegraphics*[width=18cm, height=10cm, angle=0]{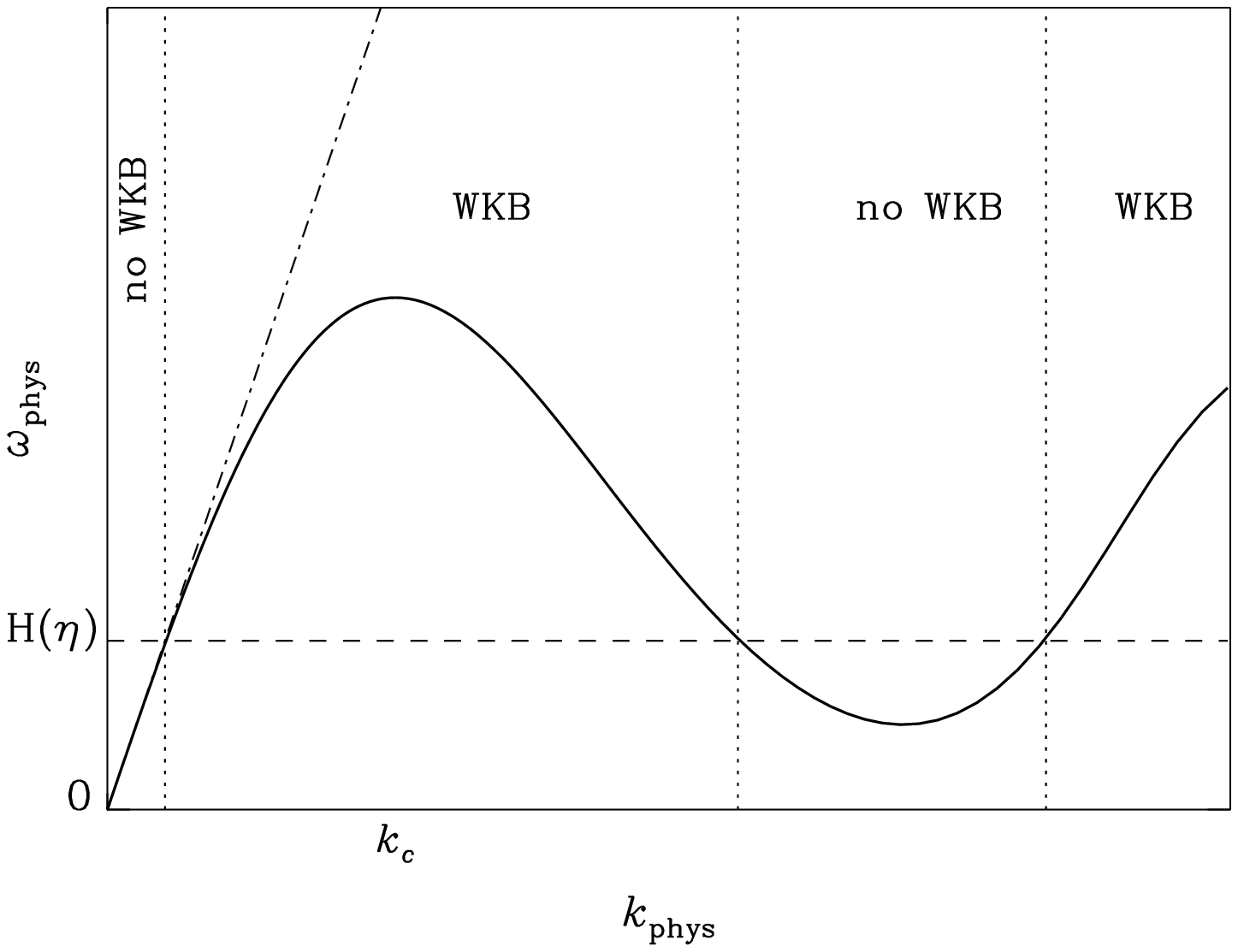} 
\caption{Typical example of a dispersion relation which breaks the  
WKB approximation in the trans-Planckian regime but allows a 
non-ambiguous  definition of the initial state. } 
\label{disp} 
\end{figure*}

The main shortcomings of describing the trans-Planckian physics by a 
modified dispersion relation is that we need to assume something about 
the physics beyond the Planck, contrary to the kind of modification 
envisaged above. Modifications in the power spectrum are obtained if 
the WKB approximation is violated in the trans-Planckian regime. A 
typical example where this happens is for the dispersion relation 
$\omega _{_{\rm phys}}^2=k_{_{\rm phys}}^2 - 2b_{11}k_{_{\rm 
phys}}^4 + 2b_{12}k_{_{\rm phys}}^6$, which represents the first terms 
of a systematic Taylor expansion. This dispersion relation is 
represented in Fig.~\ref{disp}. 
 
\par 
 
The equation of motion for the mode function of a scalar field  
is now given by 
\begin{equation} 
\mu ''+\biggl[\omega ^2(k,\eta )-\frac{a''}{a}\biggr]\mu=0\, , 
\end{equation} 
where $\omega $ is the comoving frequency given by $\omega (k,\eta 
)=a(\eta ) \omega _{_{\rm phys}}[k/a(\eta )]$. The difference  
compared to the 
previous section is that the equation determining the behavior of the 
mode function is modified. The mode functions depend on whether 
$w^2>a''/a$ or $w^2<a''/a$.  
 
Let us consider an expanding cosmological background. There will be
four time intervals (``regions''), see Fig.~\ref{piece}. As the
Universe expands, the physical wavenumber will decrease. The evolution
starts in Region I, where the WKB approximation is valid, the term
$w^2$ dominates and the mode function is given by
\begin{equation} 
\mu _{_{\rm I}}(\eta )=\frac{1}{\sqrt{2\omega (k,\eta )}}{\rm 
e}^{-i\int _{\eta _{\rm i}}^{\eta } \omega (k,\tau ){\rm d}\tau }\, . 
\end{equation} 
We have chosen the WKB vacuum, i.e. we have retained only one branch  
of the most general solution. At a time $\eta_1(k)$, the mode 
enters Region II, where the WKB approximation  
is violated, and the mode function can be expressed as 
\begin{equation} 
\mu _{_{\rm II}}(\eta )=C_+(k)a(\eta )+C_-(k)a(\eta )
\int _{\eta _{\rm i}'} 
^{\eta }\frac{{\rm d}\tau }{a^2(\tau )}\, . 
\end{equation} 
The coefficients $C_{\pm}(k)$ can be determined from joining the mode
function and its derivative at the matching point $\eta _1(k)$. After
more expansion [at the time $\eta_2(k)$], the mode leaves Region II
and enters Region III, where the WKB approximation is
restored. However, now the mode function has two branches, the second
branch having been ``generated'' in Region II. It is here that the
trans-Planckian physics modifies the standard result (which predicts
$\beta _k=0$). The mode function can be written as
\begin{eqnarray} 
\mu _{_{\rm III}}(\eta )&=&\frac{\alpha _k}{\sqrt{2\omega (k,\eta )}} 
{\rm e}^{-i\int _{\eta _{\rm i}''}^{\eta } \omega (k,\tau ){\rm d}\tau } 
\nonumber \\
& & + \frac{\beta _k}{\sqrt{2\omega (k,\eta )}} 
{\rm e}^{i\int _{\eta _{\rm i}''}^{\eta } 
\omega (k,\tau ){\rm d}\tau }\, . 
\end{eqnarray} 
The coefficients $\alpha _k$ and $\beta _k$ can be found in terms of 
$C_{\pm }(k)$ by matching the solution (and its derivative) at $\eta 
_2(k)$ as previously.  Finally, in Region IV, the WKB approximation is 
again violated. This region corresponds to the usual super-Hubble 
region. We have 
\begin{equation} 
\mu _{_{\rm IV}}(\eta )=D_+(k)a(\eta )+D_-(k)a(\eta )\int _{\eta _{\rm 
i}'''} ^{\eta }\frac{{\rm d}\tau }{a^2(\tau )}\, . 
\end{equation} 
The transition between Regions III and IV occurs at the time 
$\eta_3(k)$, the time of Hubble radius crossing.  
Our aim is to calculate the coefficients $D_+(k)$ which determines the 
spectrum of the growing mode.  
 
\par 
 
Performing the matching at $\eta _3(k)$, we obtain 
\begin{widetext} 
\begin{equation} 
\label{mu4} 
\mu _{_{\rm IV}}(\eta )=\frac{1}{\sqrt{2\omega _3}}\biggl(\alpha _k 
{\rm e}^{-i\Omega _3}+\beta _k {\rm e}^{i\Omega _3}\biggr)
\frac{a(\eta )}{a_3} 
+\frac{1}{\sqrt{2\omega _3}}a(\eta )a_3 
\biggl(\alpha _k\gamma _3 
{\rm e}^{-i\Omega _3}+\beta _k\gamma _3^*{\rm e}^{i\Omega _3}\biggr) 
\int _{\eta _3}^{\eta }\frac{{\rm d}\tau }{a^2(\tau )}\, , 
\end{equation} 
where the subscript ``$3$'' means that the corresponding quantity is 
evaluated at $\eta _3(k)$. In the previous equation, we have 
introduced the notation $\gamma _k\equiv \omega '/(2\omega )+i\omega 
+{\cal H}$ and $\Omega _3\equiv \int _{\eta _{\rm i}''}^{\eta 
_3}\omega (k,\tau ){\rm d}\tau $. On the other hand, a lengthy but 
straightforward calculation shows that the coefficients $\alpha _k$ 
and $\beta _k$ are given by 
\begin{eqnarray} 
\alpha _k &=& \frac{i}{\sqrt{4\omega 
_1\omega _2}} \biggl[\frac{a_2}{a_1}\gamma _2^*-\frac{a_1}{a_2}\gamma 
_1 -a_1a_2\gamma _1\gamma _2^*\int _{\eta _1}^{\eta _2}\frac{{\rm 
d}\tau }{a^2(\tau )}\biggr]{\rm e}^{i(\Omega _2-\Omega _1)}\, , 
\\ 
\beta _k &=& -\frac{i}{\sqrt{4\omega 
_1\omega _2}} \biggl[\frac{a_2}{a_1}\gamma _2-\frac{a_1}{a_2}\gamma 
_1 -a_1a_2\gamma _1\gamma _2\int _{\eta _1}^{\eta _2}\frac{{\rm 
d}\tau }{a^2(\tau )}\biggr]{\rm e}^{-i(\Omega _2+\Omega _1)}\, . 
\end{eqnarray} 
These expressions are similar to those found in Refs.~\cite{LLMU} and 
\cite{BJM}. So far, no approximation has been made. To go further, one 
has to take into account the fact that the difference $\vert \eta 
_2(k)-\eta _1(k)\vert$ cannot be too large. Otherwise, this would mean 
that particles production in region III is too important and, as a 
consequence, that the calculation is not valid due to this 
back-reaction problem. Therefore, we can write $\eta _1=\eta 
_2(1+\Delta )$ and perform an expansion of the coefficients $\alpha 
_k$ and $\beta _k$ in terms of the parameter $\Delta $. The result reads 
\begin{equation} 
\alpha _k={\rm e}^{i(\Omega _2-\Omega _1)} 
\biggl[1+\frac{i}{2}\omega _2\eta _2\biggl(1+\frac{Q_2}{\omega _2^2} 
-\frac{a_2''}{\omega _2^2a_2}\biggl)\Delta \biggr]+{\cal O}(\Delta^2 )\, , 
\quad   
\beta _k=\frac{i}{2}{\rm e}^{-i(\Omega _2+\Omega _1)} 
\omega _2\eta _2\biggl(1-\frac{Q_2}{\omega _2^2} 
+\frac{a_2''}{\omega _2^2a_2}\biggl)\Delta +{\cal O}(\Delta^2 )\, . 
\end{equation} 
In this expression, $Q$ is the parameter which controls the accuracy
of the WKB approximation. It is defined by $Q\equiv -\omega
''/(2\omega )+3(\omega ')^2/(4\omega ^2)$ \cite{MSwkb}. As expected,
if we send the parameter $\Delta $ to zero, then $\beta _k$ vanishes
and $\alpha _k$ becomes unity. Inserting the above expressions for
$\alpha _k$ and $\beta _k$ into Eq.~(\ref{mu4}), one arrives at
\begin{equation} 
\label{D+} 
\vert D_+(k)\vert ^2=\frac{1}{2a_3^2\omega _3} \biggl\{1-\omega 
_2\eta _2\biggl(1-\frac{Q_2}{\omega _2^2} +\frac{a_2''}{\omega 
_2^2a_2}\biggl)\sin \biggl[\int _{\eta _1}^{\eta _3} \omega (k,\tau 
){\rm d}\tau \biggr]\Delta +{\cal O}(\Delta ^2)\biggr\}\, . 
\end{equation} 
\begin{figure*}[t] 
\includegraphics*[width=18cm, height=10cm, angle=0]{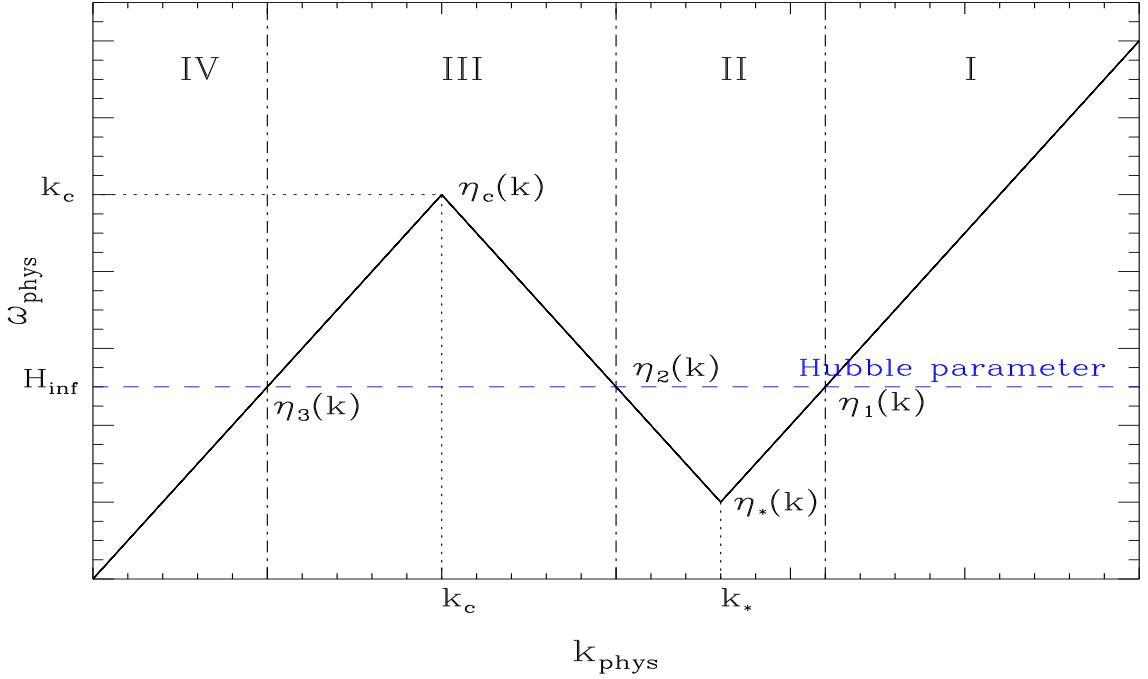} 
\caption{Approximate form of the dispersion relation considered in
Fig.~\ref{disp}. Region II is the region where the WKB approximation
is not satisfied.}
\label{piece} 
\end{figure*} 
Notice that in order to establish the previous expression we have just 
assumed that $\Delta $ is a small number but we have not assumed that 
$k_{_{\rm C}}/H$ or $k_*/H$ are small. As expected, The correction to 
$\vert D_+(k)\vert ^2$ is given by the Bogoluibov coefficient $\beta 
_k$. 
 
\par 
 
We now need to calculate explicitly the previous quantities. For this 
purpose, we introduce an approximate piecewise form of the dispersion 
relation considered before, namely 
\begin{equation} 
\omega _{_{\rm phys}}= 
\cases{ 
k_{_{\rm phys}}\, & 
$k_{_{\rm phys }}<k_{_{\rm C}} \, ,$ 
\cr 
 & \cr 
\alpha (k_{_{\rm C}}-k_{_{\rm phys }})+k_{_{\rm C}}\, &   
$k_{_{\rm C}}<k_{_{\rm phys }}<k_*\, ,$ 
\cr 
 & \cr 
sk_{_{\rm phys }}+(\alpha +1)k_{_{\rm C}}-(\alpha +s)k_*\, & 
$k_{_{\rm phys }}>k_* \, .$ } 
\end{equation} 
This piecewise dispersion relation is represented in
Fig.~\ref{piece}. The parameter $\alpha $ controls the slope of the
dispersion relation in the region where the group velocity does not
have the same sign as the phase velocity. The parameter $s$ controls
the slope in the region $k_{_{\rm phys}}>k_*$. Assuming that the
space-time is de Sitter, one can also compute the comoving frequency
$\omega =a\omega _{_{\rm phys}}(k/a)$
\begin{equation} 
\omega (k,\eta )= 
\cases{ 
k\, & 
$k_{_{\rm phys }}<k_{_{\rm C}} \, ,$ 
\cr 
 & \cr 
\displaystyle{-\alpha k-\frac{\alpha +1}{\eta }
\frac{k_{_{\rm C}}}{H}}\, &   
$k_{_{\rm C}}<k_{_{\rm phys }}<k_*\, ,$ 
\cr 
 & \cr 
\displaystyle{sk+\frac{1}{\eta }\biggl[ 
(\alpha +s)\frac{k_*}{H} 
-(\alpha +1)\frac{k_{_{\rm C}}}{H}\biggr]}\, & 
$k_{_{\rm phys }}>k_* \, .$ } 
\end{equation} 
\end{widetext}
Obviously, one has to choose $k_{_{\rm C}}>\alpha k_*/(\alpha +1)$ in 
order to insure that the frequency remains positive. The various times 
of matching can now be calculated very simply. They read 
\begin{eqnarray} 
\eta _1(k) &=& -\frac{1}{ks}\biggl[\sqrt{2}+(\alpha 
+s)\frac{k_*}{H}-(\alpha +1)\frac{k_{_{\rm C}}}{H}\biggl]\, ,  
\\ 
\eta _2(k) &=& -\frac{1}{k}\biggl(-\frac{\sqrt{2}}{\alpha 
}+\frac{\alpha +1}{\alpha }\frac{k_{_{\rm C}}}{H}\biggr)\, , 
\\ 
\eta _3(k) &=& -\frac{1}{k}\, . 
\end {eqnarray} 
The times at which $\omega _{_{\rm phys}}=k_{_{\rm C}}$ and $\omega 
_{_{\rm phys}}=k_*$, i.e. $\eta _{_{\rm C}}(k)$ and $\eta _*(k)$ 
respectively (see Fig.~\ref{piece}), can also be determined 
easily. They are given by $\eta _{_{\rm C}}(k)= -(1/k)(k_{_{\rm 
C}}/H)$ and $\eta _*(k)= -(1/k)(k_*/H)$.  
 
\par 
 
We have based the calculation of $\vert D_+(k)\vert ^2$ on the 
assumption that the parameter $\Delta $ is small in order to avoid a 
back-reaction problem. This means that the time spent by the modes of 
interest in the region where the WKB approximation is violated is 
small. In turn, this requires a link between the two scales $k_{_{\rm 
C}}$ and $k_*$ which characterizes the shape of the dispersion relation 
and the Hubble constant $H$ which characterizes the ``velocity'' with 
which a mode crosses the region where the WKB approximation is not 
valid. Using the expressions of $\eta _1(k)$ and $\eta _2(k)$, one  
finds that the link between $k_*$, $k_{_{\rm C}}$ and $\Delta $ 
can be expressed as
\begin{equation} 
\frac{k_*}{H}=\biggl[\frac{1}{\alpha }+ 
\frac{s}{\alpha (\alpha +s)}\Delta \biggr] 
\biggl[-\sqrt{2}+(\alpha +1)\frac{k_{_{\rm C}}}{H}\biggr]\, . 
\end{equation} 
If $k_{_{\rm C}}\gg H$ and $k_*\gg H$, then one has $k_*\simeq (\alpha 
+1)k_{_{\rm C}}/\alpha $, as expected. The final result can be 
expressed in terms of $k_{_{\rm C}}/H$ and $\Delta $ only. 
 
\par 
 
We can now calculate each term present in 
Eq.~(\ref{D+}). Straightforward calculations show that $Q/\omega 
^2\vert _2=-(\alpha +1)k_{_{\rm C}}/(2\sqrt{2}H)+3(\alpha +1)^2 
k_{_{\rm C}}^2/(16H^2)$, $a''/(a\omega ^2)\vert _2=1$ and $\omega \eta 
\vert _2=-\sqrt{2}$. In order to calculate the integral appearing in 
the argument of the sine function in Eq.~(\ref{D+}), one has to cut it 
into several pieces and to use the corresponding form of the piecewise 
dispersion relation. One obtains 
\begin{eqnarray} 
& & \int _{\eta _1}^{\eta _3}\omega (k,\tau ){\rm d}\tau  
=\int _{\eta _2}^{\eta _3}\omega (k,\tau ){\rm d}\tau -\sqrt{2}\Delta  
\\ 
& & =(\sqrt{2}-1)-(\alpha +1)\frac{k_{_{\rm C}}}{H}\ln  
\biggl[\frac{1}{(\alpha +1)-\sqrt{2}H/k_{_{\rm C}}}\biggr] 
\nonumber  
\\ 
& & -\sqrt{2}\Delta \, . 
\end{eqnarray} 
We have to insert this last expression into the sine function and 
expand the resulting expression once more in terms of the parameter $\Delta 
$.  The final result reads 
\begin{widetext} 
\begin{eqnarray} 
k^3P_{\chi }(k)&=& \biggl(\frac{H}{2\pi }\biggr)^2 
\biggl\{1+\sqrt{2}\biggl[2+\frac{\alpha +1}{2\sqrt{2}}\frac{k_{_{\rm 
C}}}{H} -\frac{3(\alpha +1)^2}{16}\frac{k_{_{\rm 
C}}^2}{H^2}\biggr]\sin \biggl[(\sqrt{2}-1) -(\alpha +1)\frac{k_{_{\rm 
C}}}{H} \ln \biggl(\frac{1}{\alpha +1-\sqrt{2}H/k_{_{\rm C}}}\biggr) 
\biggr]\Delta  
\nonumber \\ 
& &+{\cal O}(\Delta ^2) \biggr\}\, . 
\end{eqnarray} 
\end{widetext} 
As expected the result does not depend on $k$ since the scalar field 
lives in de Sitter space-time. It is worth noticing that we have 
not assumed anything about the ratio $k_{_{\rm C}}/H$ in order to 
obtain the previous expression (only the parameter $\Delta $ was 
supposed to be small). The previous expression is the main result of 
this section. It should be compared with Eqs.~(\ref{minkops}) and 
(\ref{dpsgw}) obtained previously. We see that the magnitude of the 
correction is no longer controled by any power of the ratio of two 
scales as it was before. The magnitude of the effect is determined by 
the time spent in the region where the WKB approximation is violated 
which in turn depends on the shape of the dispersion relation that was 
assumed.  
 
\par 
 
Finally, let us return to the back-reaction problem.  There is 
no back-reaction problem if $\vert \beta _k\vert \ll 1$. Using the 
expression of $\beta _k$ derived before and assuming that $k_{_{\rm 
C}}/H\gg 1$, we see that the parameter $\Delta $ must satisfy $\Delta 
\ll H^2/k_{_{\rm C}}^2$. This amounts to a severe fine-tuning of the 
scales $k_{_{\rm C}}$, $k_*$ and of the Hubble parameter $H$ during 
inflation in order to satisfy this condition.

\section{Conclusions}

We have analyzed the magnitude of correction terms to the power
spectra of scalar metric (density) fluctuations and gravitational
waves in inflationary cosmology under the assumption that fluctuation
modes are generated when their physical length scale equals some
critical length determined by the unknown Planck-scale physics, but
without modifying the equations of motion for the fluctuations.  The
magnitude of the correction terms can then be expressed as a function
of the dimensionless ratio $\sigma_0 = H_0 / M_{_{\rm C}}$, where
$H_0$ is the characteristic Hubble expansion rate during inflation,
and $M_{_{\rm C}}$ is the mass scale at which the new physics sets in.

\par

It is important to realize that the magnitude of the correction terms
is in general different for gravitational waves and for scalar metric
fluctuations - a point not realized in some papers on the
``trans-Planckian problem'' of inflationary cosmology.  In addition,
the magnitude of the correction terms depends sensitively on the
initial state chosen. We have shown that for the local Minkowski
vacuum state, the correction terms are of the order $\sigma_0^3$ (in
agreement with the results of \cite{npc}), whereas for nontrivial
$\alpha$-vacua the effects are much larger. If the Bogoliubov
coefficients which describe the mode mixing do not depend on
$\sigma_0$, then the correction terms can be of order unity. In the
case of Danielsson's $\alpha$-vacuum for which the Bogoliubov
coefficients depend on $\sigma_0$, the corrections to the
gravitational wave spectrum are suppressed by one power of $\sigma_0$
(in agreement with the results of \cite{Dan1}), the corrections to the
scalar metric fluctuation spectrum by two powers of $\sigma_0$.  Thus,
the consistency relation for fluctuations in inflationary cosmology
obtains corrections of linear order in $\sigma_0$, as already
emphasized in \cite{Hui}.

\par

In the final section of the paper we compared the results
obtained in earlier sections with the results obtained by
assuming that trans-Planckian physics leads to a modified
dispersion relation. In this case, corrections to the usual
power spectra of fluctuations can be obtained which are
not suppressed by any small dimensionless combinations
of energy scales in the problem. However, demanding that
the back-reaction remains under control leads to severe
fine-tuning requirements on such models.
  
\acknowledgments 
 
The work of R.~B. is supported in part by the U.S. Department of Energy under
Contract DE-FG02-91ER40688, TASK A.

\end{document}